%% file: main.tex
\documentclass[11pt,a4paper]{./lhctopnote} 
\usepackage{./lhctopphysics}
\usepackage{subfigure}
\usepackage{mathrsfs}
\usepackage{cite}
\usepackage{xspace}
\usepackage{graphicx}
\usepackage{multirow}
\usepackage{url}
\usepackage{hyperref}
\usepackage{verbatim}
\usepackage{rotating}
\graphicspath{{f/}}
\DeclareGraphicsExtensions{.pdf}

\input{newco}

\skipbeforetitle{100pt}

\title{First combination of Tevatron and LHC measurements of the top-quark mass}

\usepackage{authblk}
\author[a] {The ATLAS, CDF, CMS and D0 Collaborations\footnote{
    Work within the Tevatron Electroweak
    (TEV-EW-WG) and the Top Physics LHC (TOP-LHC-WG) working groups. \newline More
    information at http://tevewwg.fnal.gov and http://twiki.cern.ch/twiki/bin/view/LHCPhysics/TopLHCWG.}}

\atlasnote{ATLAS-CONF-2014-008\\ CDF Note 11071\\ CMS PAS
  TOP-13-014\\ D0 Note 6416 }

\date{March 17, 2014}
\abstracttext{We present a combination of measurements of the mass of
  the top quark, \mt, performed by the CDF and D0 experiments at the
  Tevatron collider and the ATLAS and CMS experiments at the Large
  Hadron Collider (LHC). The Tevatron
  data correspond to an integrated luminosity of up to 8.7~fb${^{-1}}$
  of proton-antiproton collisions from Run~II of the Tevatron at a
  centre-of-mass energy of 1.96~TeV. The LHC data correspond to an
  integrated luminosity of up to 4.9~fb${^{-1}}$ of proton-proton
  collisions from the run at a centre-of-mass energy of 7~\TeV. The
  combination includes measurements in the \ttbarlj, \ttbarll,
  \ttbaraj and \ttbarmet\ final states.  The resulting combined
  measurement of \mt\ is $173.34 \pm 0.27 \mbox{\thinspace (stat)}
  \pm 0.71 \mbox{\thinspace (syst)}$ GeV, with a total uncertainty of
  $0.76$~GeV.}

\begin{document}

\input{document.tex}

\appendix
\input{appendix.tex}

\clearpage
\bibliographystyle{./atlasBibStyleWithTitle.bst}

\bibliography{LHCmtop.bib}

\end{document}

%% file: newco.tex
\newcommand{\tevlum}{\ensuremath{8.7}\xspace}

\newcommand{\ljets}{\ensuremath{\mbox{{\em l}+jets}}\xspace}
\newcommand{\alljets}{\ensuremath{\mbox{all~jets}}\xspace}

\newcommand{\dil}{\ensuremath{\mbox{di-{\em l}}}\xspace}
\newcommand{\dilepton}{\ensuremath{\mbox{dilepton}}\xspace}
\newcommand{\metjets}{\ensuremath{E_{\rm T}^{\rm miss}\mbox{+jets}}\xspace}
\newcommand{\mt     }{\ensuremath{m_{\mathrm{top}}}\xspace}

\newcommand{\ttbarll}{\ensuremath{\ttbar\to\mbox{dilepton}}\xspace}
\newcommand{\ttbarlj}{\ensuremath{\ttbar\to\mbox{lepton+jets}}\xspace}
\newcommand{\ttbaraj}{\ensuremath{\ttbar\to\mbox{all jets}}\xspace}
\newcommand{\ttbarmet}{\ensuremath{\ttbar\to E_{\rm T}^{\rm miss}\mbox{+jets}}\xspace}

\newcommand{\lumi   }{\ensuremath{{\mathcal L}}\xspace}

\newcommand{\ppbar}{\ensuremath{p\overline{p}}}

\newcommand{\Pythia}{{\sc Pythia}}
\newcommand{\Herwig}{{\sc Herwig}}
\newcommand{\Alpgen}{{\sc Alpgen}}
\newcommand{\Powheg}{{\sc Powheg}}
\newcommand{\Madgraph}{{\sc MadGraph}}
\newcommand{\MCATNLO}{{\sc MC@NLO}}

%% file: document.tex
\section{Introduction}
The mass of the top quark (\mt) is an important parameter of the standard
model of particle physics (SM).
Precise measurements of \mt\ provide critical inputs to fits of global
electroweak parameters~\cite{LEPewkfits, GFitter} that constrain the
properties of the Higgs boson, and help assess the internal
consistency of the SM and of its extensions. In addition, the value of
\mt\ affects the stability of the SM Higgs potential, which has
 cosmological implications~\cite{HiggsStab, HiggsStab1, HiggsStab2}.

Many measurements of \mt\ have been performed by the CDF and D0
collaborations based on Tevatron proton-antiproton (\ppbar) data from Run~I (at a
centre-of-mass energy, $\sqrt{s}$, of 1.8~\TeV) and Run~II ($\sqrt{s}
= 1.96~\TeV$), corresponding to integrated luminosities
($\lumi_{\rm{int}}$) of up to \tevlum~fb$^{-1}$.
In addition, measurements of \mt\ from the LHC by ATLAS and CMS,
based on proton-proton ($pp$) collisions at 
$\sqrt{s} = 7~\TeV$, recorded during 2010 and 2011 for integrated
luminosities of up to $4.9$~fb$^{-1}$, have become available.

The Tevatron \mt\ combination results in
$\mt=173.20 \pm 0.51 \mbox{\thinspace (stat)} \pm 0.71 \mbox{\thinspace (syst)} \GeV\equiv
173.20 \pm 0.87$~\GeV~\cite{TEV2013}. The
corresponding LHC combination yields $\mt=173.29 \pm 0.23 \mbox{\thinspace 
  (stat)} \pm 0.92 \mbox{\thinspace  (syst)} \GeV \equiv 173.29 \pm
0.95$~\GeV~\cite{LHC2013}.  
This analysis combines the most
precise individual \mt\ results in each \ttbar\ final state, for each
collaboration, to get the
best overall estimate.

This note describes the first combination of Tevatron and LHC \mt\
measurements, 
six from the Tevatron collider,
based on Run II \ppbar\ data collected at $\sqrt{s}=1.96~\TeV$,
and five from the LHC, based on the $pp$ data at
$\sqrt{s}=7~\TeV$.
For CDF, these measurements include the \mt results obtained for the
\ttbarlj, \ttbarll, \ttbaraj, and the \ttbarmet\ channels\footnote{The
  \ttbarlj, \ttbarll, and \ttbaraj\ channels correspond to the
  experimental final states related to the \ttbar\ decays $\ttbar \to
  \bar l \nu b~q\bar q' \bar b$, $\ttbar \to \bar l \nu b~ l \bar\nu \bar
  b$ and $\ttbar \to q\bar q' b~q\bar q' \bar b$, respectively. The
  \ttbarmet\ channel relates to the final state from $\ttbar \to
  \bar l \nu b~q\bar q' \bar b$ selected using missing transverse energy
  signatures, \met, rather than explicit charged lepton identification
  criteria.} using up to $\lumi_{\rm{int}}=8.7$~fb$^{-1}$ of
data~\cite{CDFlj2012, CDFdil2011,
  CDFaj2012, CDFmet2013}.
For D0, the measurements are for the
\ttbarlj, and \ttbarll\ channels using up to
$\lumi_{\rm{int}}=5.3$~fb$^{-1}$ of data~\cite{D0lj2011, D0dil2012}.
The ATLAS measurements comprise the results obtained in the
\ttbarlj\ and the \ttbarll\ channels using 
$\lumi_{\rm{int}}=4.7$~fb$^{-1}$ of data~\cite{ATLlj2011, ATLdil2011}.
For CMS, the measurements, based on up to $\lumi_{\rm{int}} =
4.9$~fb$^{-1}$ of data, refer to the \ttbarlj, \ttbarll and \ttbaraj\
channels~\cite{CMSlj2011, CMSdil2011, CMSaj2011}.
In all measurements considered in the present combination,
the 
analyses are calibrated to the Monte Carlo (MC) top-quark mass
definition. 
It is expected that the difference between the MC mass definition and
the formal pole mass of the top quark is up to the order of 1~\GeV\ (see Refs.
\cite{MassDef,MassReview2013} and references therein).

This document is organised as follows.  After a brief description in
Section~\ref{sec:blue} of the methodology used for the combination, an
overview of the input measurements is given in
Sections~\ref{sec:calib} and \ref{sec:input}.  Details of the mapping
between the categories of uncertainties for CDF, D0, ATLAS and CMS,
and their corresponding correlations, are described in
Section~\ref{sec:syst}. The results of the combination are presented
in Section~\ref{sec:results}, followed in Section~\ref{sec:comments}
by a discussion of their dependence on the categorisation of the
uncertainties and on the assumed correlations.  Finally, the summary
and conclusions are given in Section~\ref{sec:summary}.


\section{Methodology}
\label{sec:blue}

The combination is performed using the Best Linear Unbiased Estimate
(BLUE) method~\cite{BLUE1, BLUE2}, implemented as described in
Ref.~\cite{BLUEcpp}.  BLUE determines the coefficients (weights) to be
used in a linear combination of the input measurements by minimising
the total uncertainty of the combined result.  In the algorithm,
assuming that all uncertainties are distributed according to Gaussian
probability density functions, both statistical and systematic
uncertainties, and their correlations, are taken into account.
A realistic estimate of the correlations is made and the effect
of the various assumptions on the final result is evaluated.

\section{Measurements calibration using Monte Carlo simulation}
\label{sec:calib}

The MC generation of signal \ttbar\ and background events proceeds
through the simulation of a primary hard interaction process ({\em
  e.g.}, $q\bar q, gg\to \ttbar$), accompanied by parton showers, and
by non-perturbative interactions that convert the showers into
colourless hadrons. In a subsequent step, soft interactions
reflected in the underlying event are also included in the
calculations~\cite{MassDef}. The simulated events are processed
through experiment-specific simulation and reconstruction software,
and the reconstructed final state particles are clustered into jets,
that can be associated with the initial partons. Within CDF, jets are
reconstructed using a cone algorithm with radius parameter
$R=0.4$~\cite{CDFcone}, while D0 employs a midpoint iterative
seed-based cone algorithm, with $R=0.5$~\cite{D0cone}. In both cases,
calorimeter information is used as input of the clustering algorithms
and $R$ is defined as $\sqrt{\Delta \eta^2 +\Delta \phi^2}$, where
$\Delta \eta$ is the pseudo-rapidity and $\Delta \phi$ is the azimuthal
angle of a given calorimeter energy deposit relative to the direction
of the jet. Within ATLAS, starting from energy clusters of adjacent
calorimeter cells called topological clusters~\cite{ATLcluster,
  ATLASJESPAPER2010, ATLASJES}, jets are formed using the anti-$k_t$
algorithm~\cite{ATLcone} with a radius parameter $R = 0.4$. In CMS,
events are reconstructed with the particle-flow algorithm~\cite{PFLOW}
that combines the information from all CMS sub-detectors to identify
and reconstruct individual objects produced in the $pp$ collision.
Particle flow objects are used as input for jet clustering also based
on the anti-$k_t$ algorithm, but with a distance parameter of
$R=0.5$~\cite{ATLcone}.

Jet energy scale (JES) calibration procedures, applied after jet
reconstruction, are meant to ensure the correct measurement of the
average jet energy across the whole detector, and they are designed to be
independent of additional events produced in $\ppbar$ or $pp$ collisions
(``pile-up''\footnote{Pile-up is the term given to the extra signal
  produced in the detector by $\ppbar$ or $pp$ interactions other than the primary
  hard scattering.}), especially at high luminosity regimes,
compounding on the event of interest.
In general, energy contributions from pile-up events are subtracted,
and the jet response in different regions of the calorimeter
(central-forward) is inter-calibrated.
Jet energy corrections account for the energy lost in uninstrumented
regions between calorimeter modules, for differences between
electromagnetically and hadronically interacting particles, as well as
for calorimeter module irregularities. In addition corrections for
shower particles scattered in or out of the reconstructed jets are
typically included.
The calibration procedures use single hadron calorimeter response
measurements, systematic MC simulation variations as well as in situ
techniques, where the jet transverse momentum ($p_{\rm T}$) is compared to the
$p_{\rm T}$ of a reference object.
CDF calibrates the jet transverse momentum using test-beam data and
single-particle simulated events and corrects the jet energy to the
parton level~\cite{JET-ENERGY-SCALE-CDF}. 
D0 determines the jet energy scale using photon+jets and dijet events and
calibrates jets in data as well as in MC to the observed particle
level. MC particle level jets are clustered from stable particles after
fragmentation~\cite{JET-ENERGY-SCALE-DZERO,JET-ENERGY-SCALE-DZERO-NEW}.
Similar procedures are in place within the ATLAS and CMS
collaborations~\cite{ATLASJES, ATLASJESPAPER2010, CMSJES, CMSJES2}:
after pile-up jet energy offset corrections, the reconstructed
jet energies in MC are restored to that from stable particle
jets. Residual calibrations, derived using in situ methods
where the jet transverse momentum is compared to the $p_{T}$ of a
reference object ({\it e.g.} using $\gamma/Z$+jet events), are then
applied to data relative to the MC.

Depending on the experiment, different MC programs and settings are
used in \mt\ analyses.
The baseline MC program for the simulation and calibration of the
top-quark mass analyses in the CDF experiment is
\Pythia~\cite{MCATLAS2} with Tune A~\cite{MCTUNEA}, based on CDF
underlying event data.  \Pythia\ is used for the simulation of the
hard process (using leading order, LO, matrix elements), the parton
shower and the underlying event modelling.
D0 adopts the tree-level multi-leg generator \Alpgen~\cite{D0ALPGEN68}
interfaced with \Pythia\ for parton showering, using a modified version
of Tune A~\cite{CDF-D0-Combo2012}. \Alpgen\ implements a parton-jet
matching, using the MLM prescription~\cite{MLM}, that avoids double counting of partons in
the hard process and in the parton shower in overlapping regions of jet
kinematics.
The baseline MC program for ATLAS is \Powheg\ interfaced with
\Pythia~\cite{MCATLAS,MCATLAS2}, with the Perugia 2011C
tune~\cite{MCATLASTUNE}. \Powheg\ uses next-to-leading order (NLO)
matrix element calculations interfaced with parton showers. The
\Madgraph\ tree-level multi-leg generator~\cite{MCCMS}, interfaced to
\Pythia\ with the Z2 tune~\cite{Z1TUNE,Z2TUNE}, is used within
CMS. Similarly to D0, the parton configurations at CMS generated
with \Madgraph\ are matched to parton showers~\cite{MLM}.
The parton distribution functions (PDF) CTEQ5L~\cite{CTEQ5},
CTEQ6L1~\cite{CTEQ6}, CT10~\cite{CT10} and CTEQ6.6L~\cite{CTEQ66} are
used respectively by the CDF, D0, ATLAS and CMS collaborations as
input for the matrix element calculations.

The baseline \ttbar\ signal MC used in the \mt\ analyses considered
here, and their main settings across the various experiments, are
summarised in Table~\ref{tab:MCsetup}.

\begin{table}[!t]
\begin{center}
\footnotesize 
\begin{tabular}{|r|c|c|c|c|}
\hline

Experiment & Matrix Element & Parton shower / hadronisation  & PDF & Tune \\
\hline
CDF        & \Pythia         & \Pythia                         &
CTEQ5L & Tune A\\
D0         & \Alpgen         & \Pythia                         & CTEQ6L1& Mod. Tune A\\
ATLAS      & \Powheg         & \Pythia                         & CT10 & Perugia2011C\\
CMS        & \Madgraph       & \Pythia                         & CTEQ6.6L& Z2\\
\hline
\end{tabular}
\caption{Baseline \ttbar\ signal MC used in \mt\ analyses and their
  main settings for the various experiments. See text for details
  and references.
}
\label{tab:MCsetup}

\end{center}
\end{table}

A systematic uncertainty covering differences between generator
models is assigned to the input measurements, and ranges from
0.02~\GeV\ to 0.25~\GeV, depending on the analysis (typically
$0.10$~\GeV\ for CDF, and $0.25$~\GeV\ for D0 measurements;
$0.20$~\GeV\ for ATLAS, and in the range $[0.02,~0.19]$ \GeV\ for CMS
inputs).
This is included together with other effects in the signal modelling
systematic uncertainty (labelled MC in Section~\ref{sec:syst})
described below. 
The systematic uncertainty related to the specific MC choice is found
to be marginal with respect to the possible intrinsic difference
between the top-quark mass implemented in any MC and the pole mass
definition~\cite{MassDef}.

\section{Input measurements}
\label{sec:input}

This \mt\ combination takes as inputs all of the measurements from the
previous LHC combination~\cite{LHC2013}, and a partial set of
those from the latest Tevatron combination~\cite{TEV2013}.
The chosen inputs correspond to the best measurements per channel and
per experiment (excluding results from Tevatron Run~I).
These comprise (i) four CDF published results from the \ttbarlj,
\ttbarll, \ttbaraj, and \ttbarmet\ channels~\cite{CDFlj2012,
  CDFdil2011, CDFaj2012, CDFmet2013}, (ii) two D0 published
measurements from the \ttbarlj, and \ttbarll channels~\cite{D0lj2011,
  D0dil2012}, (iii) two preliminary ATLAS results in the \ttbarlj and
\ttbarll channels~\cite{ATLlj2011,ATLdil2011}, and (iv) three
published results from the CMS collaboration in the \ttbarlj,
\ttbarll, and \ttbaraj\ channels~\cite{CMSlj2011, CMSdil2011,
  CMSaj2011}.

An overview of the input \mt\ measurements used in this combination
is shown in Table~\ref{tab:inputs}. Further details are provided in
the following sections. 

\begin{table}[!t]
\begin{center}
\footnotesize 
\begin{tabular}{|r|r|c|c|c|c|}
\hline

Experiment & \ttbar\ final state & $\lumi_{\rm{int}}$ [fb$^{-1}$] & 
\mt\ $\pm$ (stat.) $\pm$ (syst.) [\GeV] & Total uncertainty on \mt
[\GeV] ([\%]) & Reference \\

\hline 
\multirow{4}[1]{*}{CDF}   & \ljets  & 8.7 & $172.85\pm 0.52 \pm 0.99$ & 1.12 ~~~~~~(0.65) & \cite{CDFlj2012} \\
   & \dilepton  & 5.6 & $170.28\pm 1.95 \pm 3.13$ &  3.69 ~~~~~~(2.17) & \cite{CDFdil2011} \\
   & \alljets & 5.8 & $172.47\pm 1.43 \pm 1.41$ &  2.01 ~~~~~~(1.16) & \cite{CDFaj2012} \\
   & \metjets & 8.7 & $173.93\pm 1.26 \pm 1.36$ &  1.85 ~~~~~~(1.07) & \cite{CDFmet2013} \\
\hline

\multirow{2}[1]{*}{D0}    & \ljets  & 3.6 & $174.94\pm 0.83 \pm 1.25$
&  1.50 ~~~~~~(0.86) & \cite{D0lj2011} \\
    & \dilepton  & 5.3 & $174.00\pm 2.36 \pm 1.49$ &  2.79 ~~~~~~(1.60) & \cite{D0dil2012} \\

\hline

\multirow{2}[1]{*}{ATLAS}   & \ljets  & 4.7 & $172.31\pm 0.23 \pm
1.53$ &  1.55 ~~~~~~(0.90) & \cite{ATLlj2011} \\
   & \dilepton  & 4.7 & $173.09\pm 0.64 \pm 1.50$ &  1.63 ~~~~~~(0.94) & \cite{ATLdil2011} \\
\hline
\multirow{3}[1]{*}{CMS}   & \ljets  & 4.9 & $173.49\pm 0.27 \pm 1.03$
&  1.06 ~~~~~~(0.61) & \cite{CMSlj2011} \\
   & \dilepton  & 4.9 & $172.50\pm 0.43 \pm 1.46$ &  1.52 ~~~~~~(0.88) & \cite{CMSdil2011} \\
   & \alljets  & 3.5 & $173.49\pm 0.69 \pm 1.23$ &  1.41 ~~~~~~(0.81) & \cite{CMSaj2011} \\
\hline

\end{tabular}
\caption{Overview of the 11 input measurements used in this \mt\
  combination. 
}
\label{tab:inputs}

\end{center}
\end{table}

\subsection{CDF measurements}

The CDF measurements in the \ttbarlj and \ttbarmet channels are based
on the full Run~II data set of 8.7~fb$^{-1}$~\cite{CDFlj2012,
  CDFmet2013}. The \mt\ results in the \ttbarll and \ttbaraj channels
use 5.6~fb$^{-1}$ and 5.8~fb$^{-1}$ of data,
respectively~\cite{CDFdil2011, CDFaj2012}. The CDF Run~I measurements
have relatively large uncertainties and for simplicity are thus not considered in this
combination. The CDF analyses based upon charged particle tracking
that use the transverse decay length of $b$-tagged jets ($L_{\rm xy}$)
or the transverse momentum of electrons and muons from $W$ boson
decays ($p_{\rm T}^{\rm lep}$) use only part of the available Run II
data~\cite{CDFtr2010, CDFlp2011}.  
Due to their large total uncertainties and statistical correlation
with \ttbarlj and \ttbarll events, these results are not included in
this combination.

In all four CDF analyses, the template method is used, and the event
reconstruction is based on a kinematic fit to the \ttbar\ decay
hypothesis.
For example, the templates may be the top-quark mass reconstructed
from a kinematic fit in MC samples, generated using different input
\mt.  The templates are transformed to continuous functions of \mt,
either through a non-parametric kernel-density estimator~\cite{KDE},
or by fitting an analytic function that interpolates between the
discrete input values of \mt.  These  are then used in a
maximum likelihood fit to the data.

The analysis in the \ttbarll\ channel measures \mt\ using the
``neutrino weighting'' algorithm~\cite{NuWeight1, NuWeight2}.  This
procedure steps through different hypotheses for the pseudo-rapidity
distributions of the two neutrinos in the final state.  For each
hypothesis, the algorithm calculates the full event kinematics and
assigns a weight to the resulting reconstructed top-quark mass based
on the agreement between the calculated and measured missing
transverse energy. The solution corresponding to the maximum weight is
selected to represent the event.  The analysis also uses template
distributions of $m_{T2}$, a
variable related to the transverse masses of the top
quarks~\cite{CDFdil2011}.

In the case of the \ttbarlj, \ttbaraj, and \ttbarmet channels, two- or
three-dimensional template fits (depending on the number of input
template distributions utilised) are performed to determine \mt\ along
with a global jet-energy scale factor JSF (denoted as ``JES'' in the
original publications). The JSF is constrained by the response of
light-quark jets by the kinematic information in $W\to q\bar q'$
decays (referred to as in situ \ttbar\ jet energy calibration).  This
technique was pioneered in the \ttbarlj analyses by CDF and D0 at the
beginning of Run~II of the Tevatron~\cite{CDFlj2006,D0lj2006}.
In the fitting procedures, the external information about the
uncertainty on the JES is used as a prior in determining the JSF. The
resulting correlation among different CDF measurements and categories
of uncertainty is evaluated by comparing the \mt\ values both with and
without the JES priors, and found to be negligible. The jet
energy calibrations for the \ttbarlj analysis are improved using an
artificial neural network to achieve a better $b$-jet energy
resolution. In a way similar to what is described in
Ref.~\cite{CDFJetNNCorr}, this algorithm incorporates precision
tracking and secondary vertex information, in addition to standard
calorimeter measurements.

\subsection{D0 measurements}

The two D0 measurements of \mt\ used in this combination
correspond to the best D0 measurements in the \ttbarlj\ and 
\ttbarll\ channels~\cite{D0lj2011,D0dil2012}. 
D0 results from Run~I also have relatively large uncertainties, and
for simplicity are thus not used in this combination.
The  \ttbarlj\ measurement is based on 3.6~fb$^{-1}$ of Run~II
Tevatron data~\cite{D0lj2011}.  It uses a matrix element
method~\cite{D0MatMeth} with an in situ jet energy calibration.
To optimise the precision, it incorporates the constraint from the
invariant mass of the hadronically decaying $W$ boson from the top
quark ($t\to Wb$), together with an external prior on the jet energy
calibrated through studies of exclusive $\gamma$+jet and jet events.
A flavor-dependent jet response correction is further applied to MC
events~\cite{JET-ENERGY-SCALE-DZERO-NEW}. The result using
2.6~fb$^{-1}$~\cite{D0lj2011} of data is combined with the 1~fb$^{-1}$
measurement~\cite{D0lj2008} which uses statistically independent data.
To take into account correlations among different sources of
systematic uncertainty, the contribution from the JES prior is kept
separate following the procedure detailed in Ref.~\cite{D0Combo2009}.

The  \ttbarll\ measurement is based on 5.3~fb$^{-1}$ of Run~II Tevatron
data~\cite{D0dil2012}.  The measurement uses the neutrino weighting
technique, as described for the corresponding CDF analysis.
In addition, the JSF re-calibration from the  \ttbarlj\
analysis is applied to this channel, along with an estimate of the
uncertainty in transferring that calibration to the dilepton event
topology.
The result using 4.3~fb$^{-1}$~\cite{D0dil2012} of data is combined
with the 1~fb$^{-1}$ measurement~\cite{D0dil2009}. As in the case of
the \ttbarlj\ result, these analyses use statistically independent
data.

\subsection{ATLAS measurements}

All the ATLAS measurements of \mt\  rely on the template method, and
use analytic probability density functions for interpolation.

In the \ttbarlj analysis, events are reconstructed using a kinematic
fit to the \ttbar\ decay hypothesis ($\ttbar \to \bar l\nu b~q\bar q'
\bar b$). A
three-dimensional template method is used, where \mt is determined
simultaneously with a JSF from $W\to q\bar q'$ decays and a separate
$b$-to-light-quark energy scale factor (bJSF)~\cite{ATLlj2011}.
The JSF and bJSF account for differences between data and simulation
in the light-quark and in the relative $b$- and light-quark jet energy
scale, respectively, thereby mitigating the corresponding systematic
uncertainties.
No prior knowledge of the uncertainty related to the light- and
$b$-quark jet energy scales is used when determining the JSF and the
bJSF parameters.

The $\ttbarll$ analysis is based on a one-dimensional template method,
where the templates are constructed for the $m_{lb}$ observable,
defined as the per-event average invariant mass of the two lepton
(either electron or muon) plus $b$-jet pairs in each event from the
decay of the top quarks~\cite{ATLdil2011}.

\subsection{CMS measurements}  
The CMS input measurements in the \ttbarlj ~\cite{CMSlj2011} and
\ttbaraj ~\cite{CMSaj2011} channels are based on the ideogram
method~\cite{IdeogramMeth}, and employ a kinematic fit of the decay
products to a \ttbar\ hypothesis ($\ttbar \to \bar l\nu b~q\bar q'\bar b$ or $\ttbar
\to q\bar q' b~q\bar q' \bar b$). MC-based likelihood functions are exploited for each event (ideograms) that
depend only on the top-quark mass or on both \mt\ and a
JSF. The ideograms reflect the compatibility of the
kinematics of the event with a given decay hypothesis.
For the \ttbarlj analysis \mt is derived simultaneously with a JSF
from $t\to Wb~(W\to q\bar q')$ decays (two-dimensional ideogram method);
whereas for the \ttbaraj analysis only \mt is extracted from a fit to
the data (one-dimensional ideogram method).  Similar to the ATLAS \ttbarlj
analysis, no prior knowledge of the uncertainty on the jet energy scale is
used to determine the JSF.

For the CMS \ttbarll analysis, \mt\ is obtained from an analytical
matrix-weighting technique, where the full reconstruction of the event
kinematics is done under different \mt assumptions. For each event,
the most likely \mt\ hypothesis, fulfilling \ttbar\ kinematic
constraints, is obtained by assigning weights that are based on
probability density functions for the energy of the charged lepton
taken from simulation, whic are applied in the solution of the
kinematic equations~\cite{CMSdil2011}.

Results from alternative techniques~\cite{CMSendpoint2011,CMSLxy2012},
characterised by different sensitivities to the dominant systematic
contributions, have recently become available but are not included in
the present analysis.


\section{Evaluation and categorisation of uncertainties}
\label{sec:syst}

In addition to the statistical uncertainty, the measurements of \mt 
are subject to several sources of systematic effects.
These are subdivided using the following categories, that are
detailed in Sections~\ref{sec:JESunc}, \ref{sec:MCunc} and
~\ref{sec:otherunc}.

\begin{itemize}
\item {\bf JES:} this group of uncertainties stems from the limited
  understanding of the detector response to (and the modelling of)
  different types of jets ($b$-quark, light-quark or gluon originated
  jets).

\item {\bf Theory and modelling:} this class of uncertainties is
  related to the MC modelling of the \ttbar\ signal, and arises from
  several components. These range from the specific choice of the MC
  generator and the associated PDF, to the models used for the parton
  hadronisation, the underlying event, and the colour reconnection
  effects. In addition, variations of the settings used to regulate
  the QCD radiation accompanying the \ttbar\ production are
  considered.

\item {\bf Detector modelling, background contamination, environment:}
  these categories comprise systematic uncertainties stemming from
  detector resolution effects, reconstruction efficiencies, and the
  $b$-tagging performance in data relative to the MC. In addition, effects related to
  normalisation and differential distributions of backgrounds events, and the
  modelling of the data-taking conditions in the MC simulation relative
  to the data, are included in this category.
\end{itemize}

Systematic uncertainties on all eleven \mt\ input values are evaluated
by changing the respective quantities by $\pm 1$ standard deviation,
or by changing the \ttbar\ signal modelling parameters relative to the
default analysis. For each component of uncertainty, the observed
shift in \mt\ relative to the nominal analysis is used to determine
the corresponding top-quark mass uncertainty. The total uncertainty is
defined by the quadratic sum
of all individual contributions, {\it i.e.} neglecting possible
correlations among different uncertainty classes (by construction
expected to be minimal), as well as non-linear effects on the
measured value of \mt.

Depending on the methods and experimental details, different
correlations can arise among the sources of uncertainty of the eleven input
\mt\ measurements. 
The  following details how these are treated in the evaluation of the
final results.

The CDF and D0 categorisations of uncertainties, and their
assumed correlations, are
documented in Refs.~\cite{TEV2013, CDF-D0-Combo2012}.  The
categorisations for ATLAS and CMS closely follow those of
Ref.~\cite{LHC2013} (see Appendix~\ref{app:naming} for details on
naming conventions). 
In certain cases, without altering the total uncertainty of the input
measurements, the breakdowns into categories of systematic uncertainties
differ from the original publications\footnote{When asymmetric
  uncertainties were
  reported~\cite{CMSdil2011}, a symmetrisation procedure is applied taking the
  maximum between the absolute values of the positive and negative
  uncertainties.}. The latter typically have a coarser categorisation
and modifications were required to match the desired uncertainty
classes.
The correlation coefficients $\rho_{\rm CDF},~\rho_{\rm D0},~\rho_{\rm
  ATL}$, and $\rho_{\rm CMS}$ indicate the assumed correlation among
measurements within the same experiment (collectively referred to as
$\rho_{\rm EXP}$), while $\rho_{\rm LHC}$ and $\rho_{\rm TEV}$
indicate the correlation assumed between measurements at the LHC and
the Tevatron analyses. Correlation coefficients $\rho_{\rm
  ATL-TEV}$ and $\rho_{\rm CMS-TEV}$ stand for the correlations between
measurements from ATLAS or CMS and the Tevatron ($\rho_{\rm COL}$ as a
short hand notation), respectively.

Specific systematic uncertainties on individual \mt\ inputs, stemming
from modelling of production processes, detector response and other
effects,  can differ for many reasons.
Analysis-specific issues, such as the amount of kinematic information
exploited in the analysis, and the level
of sophistication of the \ttbar\ reconstruction algorithms,
can influence the sensitivity of the input measurements to different 
\ttbar\ modelling systematic uncertainties.
Similarly, differences in analysis methods, for example
the possibility to simultaneously determine global jet energy scale
factors and \mt, can lead to a mitigation of the JES-related
systematic uncertainties. 
This can reduce certain signal modelling
systematics, but possibly increase 
some detector related uncertainties.
Finally, detector performance can be affected by experimental
specifications. For example, the dependence of the JES uncertainty on
jet $p_{\rm T}$ can affect the contribution of the JES component to
the uncertainty on \mt, for different cutoffs on $p_{\rm T}$, even for
analyses implementing in situ $t\to Wb, ~W\to q\bar q'$ calibrations.

\begin{table}[t!]
\scriptsize
\begin{center}
\begin{tabular}{|l|r|r|r|r|r|r|r|r|r|r|r||r|}

\cline{2-13}
  \multicolumn{1}{c}{} 
& \multicolumn{11}{|c||}{Input measurements and uncertainties in GeV} 
& \multicolumn{ 1}{c|}{} \\ 
\hline

& \multicolumn{4}{c|}{CDF} & \multicolumn{2}{c|}{D0} &
\multicolumn{2}{c|}{ATLAS} & \multicolumn{3}{c||}{CMS} & World\\  \cline{2-12}

Uncertainty & \ljets &  \dil & \alljets & \met\ & \ljets & \dil & \ljets & \dil & \ljets & \dil & \alljets & 
Combination\\
\hline
$\mt$       & 172.85 & 170.28 & 172.47 & 173.93 & 174.94 & 174.00 & 172.31 & 173.09 & 173.49 & 172.50 & 173.49 & 173.34 \\ \hline
Stat        &   0.52 &   1.95 &   1.43 &   1.26 &   0.83 &   2.36 & 0.23 &   0.64 &  0.27 &  0.43 &  0.69 &  0.27 \\ \hline
iJES       &    0.49 &   n.a. &   0.95 &  1.05 &  0.47 &  0.55 &  0.72
&  n.a. &0.33 &  n.a. &  n.a. &  0.24 \\  
stdJES     &  0.53 &  2.99 &  0.45 &  0.44 &  0.63 &  0.56 &  0.70 &  0.89 &  0.24 &  0.78 &  0.78 &  0.20 \\ 
flavourJES &  0.09 &  0.14 &  0.03 &  0.10 &  0.26 &  0.40 &  0.36 &  0.02 &  0.11 &  0.58 &  0.58 &  0.12 \\ 
bJES &  0.16 &  0.33 &  0.15 &  0.17 &  0.07 &  0.20 &  0.08 &  0.71 &  0.61 &  0.76 &  0.49 &  0.25 \\ 
\hline
MC &  0.56 &  0.36 &  0.49 &  0.48 &  0.63 &  0.50 &  0.35 &  0.64 &  0.15 &  0.06 &  0.28 &  0.38 \\ 
Rad &  0.06 &  0.22 &  0.10 &  0.28 &  0.26 &  0.30 &  0.45 &  0.37 &  0.30 &  0.58 &  0.33 &  0.21 \\ 
CR &  0.21 &  0.51 &  0.32 &  0.28 &  0.28 &  0.55 &  0.32 &  0.29 &  0.54 &  0.13 &  0.15 &  0.31 \\ 
PDF &  0.08 &  0.31 &  0.19 &  0.16 &  0.21 &  0.30 &  0.17 &  0.12 &  0.07 &  0.09 &  0.06 &  0.09 \\ 
\hline
DetMod &  $<$0.01 &  $<$0.01 &  $<$0.01 &  $<$0.01 &  0.36 &  0.50 &  0.23 &  0.22 &  0.24 &  0.18 &  0.28 &  0.10 \\ 
$b$-tag &  0.03 &  n.e. &  0.10 &  n.e. &  0.10 &  $<$0.01 &  0.81 &  0.46 &  0.12 &  0.09 &  0.06 &  0.11 \\ 
LepPt &  0.03 &  0.27 &  n.a. &  n.a. &  0.18 &  0.35 &  0.04 &  0.12 &  0.02 &  0.14 &  n.a. &  0.02 \\ 
BGMC &  0.12 &  0.24 &  n.a. &  n.a. &  0.18 &  n.a. &  n.a. &  0.14 &  0.13 &  0.05 &  n.a. &  0.10 \\ 
BGData &  0.16 &  0.14 &  0.56 &  0.15 &  0.21 &  0.20 &  0.10 &  n.a. &  n.a. &  n.a. &  0.13 &  0.07 \\ 
Meth &  0.05 &  0.12 &  0.38 &  0.21 &  0.16 &  0.51 &  0.13 &  0.07 &  0.06 &  0.40 &  0.13 &  0.05 \\ 
MHI  &  0.07 &  0.23 &  0.08 &  0.18 &  0.05 &  $<$0.01 &  0.03 &  0.01 &  0.07 &  0.11 &  0.06 &  0.04 \\ 
\hline
Total Syst &  0.99 &  3.13 &  1.41 &  1.36 &  1.25 &  1.49 &  1.53 &  1.50 &  1.03 &  1.46 &  1.23 &  0.71 \\ 
\hline
Total        &  1.12 &  3.69 &  2.01 &  1.85 &  1.50 &  2.79 &  1.55 &  1.63 &  1.06 &  1.52 &  1.41 &  0.76 \\ 
\hline

\end{tabular}

\caption{Uncertainty categories assignment for the input measurements and
  the result of the world \mt\ combination. All values are in
  \GeV. In the table, ``n.a.'' stands for not applicable; ``n.e.''
  refers to uncertainties not evaluated (see text for details). }
\label{tab:syscat}

\end{center}
\end{table}

\begin{table}[tbp!]
\begin{center}
\footnotesize 
\begin{tabular}{|r||c|c|c|c||c|c||c|c||}
\hline
          &  \multicolumn{4}{c||}{$\rho_{\rm EXP}$}  &
 \multicolumn{1}{c|}{\multirow{2}[1]{*}{$\rho_{\rm LHC}$}} &
 \multicolumn{1}{c||}{\multirow{2}[1]{*}{$\rho_{\rm TEV}$}} &   
                    \multicolumn{2}{c||}{$\rho_{\rm COL}$} \\ \cline{2-5} \cline{8-9}
          &$\rho_{\rm CDF}$   &$\rho_{\rm D0}$   &$\rho_{\rm ATL}$
          &$\rho_{\rm CMS}$   & & &
$\rho_{\rm ATL-TEV}$   &$\rho_{\rm CMS-TEV}$  \\
\hline
    Stat  &    0.0  &    0.0  &    0.0  &    0.0  &    0.0  &    0.0
    &    0.0  &    0.0 \\
\hline
    iJES  &    0.0  &    1.0  &    0.0  &    0.0  &    0.0  &    0.0  &    0.0  &    0.0 \\
    stdJES  &    1.0  &    1.0  &    1.0  &    1.0  &    0.0  &    0.0  &    0.0  &    0.0 \\
   flavourJES  &    1.0  &    1.0  &    1.0  &    1.0  &    0.0  &    0.0  &    0.0  &    0.0 \\
    bJES  &    1.0  &    1.0  &    1.0  &    1.0  &    0.5  &    1.0  &    1.0  &    0.5 \\

\hline
      MC  &    1.0  &    1.0  &    1.0  &    1.0  &    1.0  &    1.0  &    1.0  &    1.0 \\
     Rad  &    1.0  &    1.0  &    1.0  &    1.0  &    1.0  &    1.0  &    0.5  &    0.5 \\
      CR  &    1.0  &    1.0  &    1.0  &    1.0  &    1.0  &    1.0  &    1.0  &    1.0 \\
     PDF  &    1.0  &    1.0  &    1.0  &    1.0  &    1.0  &    1.0  &    0.5  &    0.5 \\
\hline
  DetMod  &    1.0  &    1.0  &    1.0  &    1.0  &    0.0  &    0.0  &    0.0  &    0.0 \\
    $b$-tag  &    1.0  &    1.0  &    1.0  &    1.0  &    0.0  &    0.0  &    0.0  &    0.0 \\
    LepPt  &    1.0  &    1.0  &    1.0  &    1.0  &    0.0  &    0.0  &    0.0  &    0.0 \\
    BGMC$^\dag$  &    1.0  &    1.0  &    1.0  &    1.0  &    1.0  &    1.0  &    1.0  &    1.0 \\
    BGData  &    0.0  &    0.0  &    0.0  &    0.0  &    0.0  &    0.0  &    0.0  &    0.0 \\
    Meth  &    0.0  &    0.0  &    0.0  &    0.0  &    0.0  &    0.0  &    0.0  &    0.0 \\
    MHI   &    1.0  &    1.0  &    1.0  &    1.0  &    1.0  &    0.0  &    0.0  &    0.0 \\
\hline
\end{tabular}
\end{center}

\caption{Assumed correlation coefficients for each source of uncertainty. 
  The symbols $\rho_{\rm CDF},~\rho_{\rm D0},~\rho_{\rm ATL}$, and $\rho_{\rm CMS}$
  represent the assumed correlations among measurements from the
  same experiment, while $\rho_{\rm LHC}$ and $\rho_{\rm TEV}$ indicate the correlations
  assumed respectively between measurements at the LHC and at the
  Tevatron. The  $\rho_{\rm ATL-TEV}$ and $\rho_{\rm CMS-TEV}$ reflect the
  correlations between measurements from ATLAS or CMS and the Tevatron. 
  \newline
  $^\dag$ For the BGMC,
  the 100\% correlation is assumed only for measurements using the same
  \ttbar\ final state. }
\label{tab:corr}
\end{table}

The uncertainties and the assumed correlations among classes are
summarised in Tables~\ref{tab:syscat}
and~\ref{tab:corr}, respectively, and detailed below.
These reflect the present understanding and the limitations
originating from independent paths followed by the experiments
to evaluate the individual sources of uncertainty.  The stability of the
result under different assumptions is discussed in
Section~\ref{sec:comments}.

\subsection{Statistical uncertainty}

\begin{itemize}

\item[Stat:] This is the statistical uncertainty associated with the
  \mt\ determination from the available data. It is 
  uncorrelated between different \ttbar\ final states, experiments and
  the two colliders (orthogonal data samples). 

\end{itemize}

\subsection{JES uncertainties}
\label{sec:JESunc}
The following systematic uncertainties stem from the limited knowledge
of the JES~\cite{JET-ENERGY-SCALE-CDF,
  JET-ENERGY-SCALE-DZERO,JET-ENERGY-SCALE-DZERO-NEW ,ATLASJES, ATLASJES2, ATLASJESPAPER2010, CMSJES, CMSJES2}. 
Since the methodologies and assumptions to derive JES
corrections and their corresponding uncertainties are not always
directly comparable between experiments, variations of the correlation
assumptions described below are considered in checking the stability of
the combination (see Section~\ref{sec:comments}).

\begin{itemize}

\item[iJES:] This is the part of the JES uncertainty of the \mt
  measurements that originates from in situ \ttbar\ ($t\to Wb, W\to
  q\bar q'$) calibration procedures. Being statistical in nature, it
  is
  uncorrelated among the individual measurements. For analyses
  performing an in situ jet calibration based on the simultaneous fit
  of the reconstructed $W$ boson and top quark invariant masses, this
  corresponds to the additional statistical uncertainty associated
  with the simultaneous determination of a JSF using the $W\to q\bar q'$
  invariant mass and \mt~\cite{CDFlj2012,CDFaj2012, CDFmet2013,D0lj2011,
    ATLlj2011, CMSlj2011}. For the ATLAS \ttbarlj\
  measurement~\cite{ATLlj2011}, it also includes the extra statistical
  component due to the simultaneous determination of a bJSF. For this
  category, we assume that uncertainties are uncorrelated ($\rho_{\rm
    CDF}=\rho_{\rm ATL}=\rho_{\rm CMS}=0$), except for the D0 \ttbarlj
  and \ttbarll measurements, where the result for the JSF from the
  \ttbarlj measurement is used to constrain the JES in the \ttbarll
  analysis ($\rho_{\rm D0}=1$).

\item[stdJES:] (Standard light jet energy scale uncertainty, dJES in
  Refs.~\cite{TEV2013,LHC2013}) This refers to the standard, non-flavour specific,  part of the JES uncertainty.
  It is assumed to be correlated between the measurements in the
  same experiment but not correlated between experiments nor across
  colliders ($\rho_{\rm EXP}= 1$, $\rho_{\rm LHC}=\rho_{\rm TEV}=0$ and $\rho_{\rm COL}
  \equiv \rho_{\rm ATL-TEV} = \rho_{\rm CMS-TEV}=0$ ).

  For CDF, this includes uncertainties on the relative jet energy
  correction as a function of jet $\eta$. This is evaluated using
  dijet data, along with \Pythia\ and \Herwig~\cite{HERWIG} simulated dijet
  samples~\cite{JET-ENERGY-SCALE-CDF}.  In addition, uncertainties
  originating from the attempt to correct the jet energy to the parton
  level are included in this category for all CDF
  measurements. These are related to the out-of-cone showering
  corrections to the MC showers (cJES in Ref.~\cite{TEV2013}), the
  absolute calibration, and the modelling of the multiple hadronic
  interaction and the underlying events (rJES in Ref.~\cite{TEV2013}).

  For D0, this uncertainty term represents almost all parts
  of JES calibrations. The absolute energy scale for jets in data is
  calibrated using $\gamma$+jet data
  using the ``{\met} projection fraction''
  method~\cite{JET-ENERGY-SCALE-DZERO}. Simulated samples of
  $\gamma$+jets and $Z$+jets events are compared to data, and used
  to derive jet energy scale corrections for MC and data events. The JES
  is also corrected as a function of $\eta$ for forward jets relative to the
  central jets using $\gamma$+jets and dijets data. Out-of-cone
  particle scattering corrections are determined with $\gamma$+jets
  simulated events.

  For LHC experiments, the stdJES uncertainty category comprises three
  components (uncorrJES, insitu$\gamma/Z$ JES, and
  intercalibJES)~\cite{LHC2013}, which in the present analysis are
  summed in quadrature. These are assumed to be fully correlated
  between measurements from the same experiment, but uncorrelated
  across ATLAS and CMS ($\rho_{\rm ATL} = \rho_{\rm CMS} =1$, and
  $\rho_{\rm LHC}=0$).

  \begin{itemize}
  \item[uncorrJES:] (LHC only, part of stdJES) For ATLAS this includes
    contributions from the limited data sample statistics used to
    derive the standard jet energy calibrations. In addition,
    uncertainty contributions from detector-specific components,
    pile-up suppression techniques, and the
    presence of close-by jet activity are included in this source.
    For CMS, this uncertainty source includes the statistical
    uncertainty of the standard jet energy calibration, contributions
    stemming from the jet energy correction due to pile-up effects,
    uncertainties due to the variations of the calorimeter response
    versus time, and detector specific effects.

  \item[insitu$\gamma/Z$JES:] (LHC only, part of stdJES) This corresponds to the part of the JES
    uncertainty stemming from modelling uncertainties affecting the
    JES determination using $\gamma/Z$+jets events, not
    included in the uncorrJES category. 

  \item[intercalibJES:] (LHC only, part of stdJES) This is the JES uncertainty component
    originating from the 
    relative jet
    $\eta$ (central-forward) and $p_{\rm T}$ inter-calibration
    procedures. Within CMS, when evaluating this uncertainty
    contribution, an extrapolation to zero radiation is performed, and
    sizable statistical contributions are present\footnote{For the sake of
    simplicity and opposite to what was done in Ref.~\cite{LHC2013},
    the combination is carried out with $\rho_{\rm ATL} = \rho_{\rm CMS}
    =1;~\rho_{\rm LHC}=0$.}.
\end{itemize}

\item[flavourJES:] This includes the part of the JES uncertainty
  stemming from differences in the jet energy response for various jet
  flavours (quark- versus gluon-originated jets) and variations of the
  flavour mixture with respect to that used in the calibration
  procedures.
  Contributions due to the modelling of $b$-quark jets are
  treated separately and discussed below.  The combined \mt\ result is
  obtained with $\rho_{\rm EXP} =1;~ \rho_{\rm TEV}=\rho_{\rm
    LHC}=\rho_{\rm COL}=0$.

\item[bJES:] This accounts for an additional $b$-jet specific
  uncertainty, arising from the uncertainty in the modelling of the
  response of jets originating from $b$-quarks~\cite{CDF-D0-Combo2012,
    ATLASJES2, CMSJES}.  

  In CDF and D0, this category covers the uncertainty on the
  semileptonic branching fraction $(10.69 \pm 0.22) \times
  10^{-2}$~\cite{PDGBook20067} of $B$~hadrons. Both collaborations
  re-weight \ttbar\ events by the uncertainty on the central value
  ($\pm 2.1$\%), and take half of the resulting mass difference as the
  uncertainty on \mt. 
  In addition, this category covers the uncertainty on the $b$-jet
  fragmentation. CDF uses the default \Pythia\ model of $b$-jet
  fragmentation based on the Bowler model~\cite{BOWLER-B-DECAY}.  D0
  uses a model with the $b$-fragmentation parameters tuned to data from ALEPH, DELPHI,
  and OPAL~\cite{SLD_BFRAG, ALEPH_BFRAG, DELPHI_BFRAG, OPAL_BFRAG,
    B-TUNING}. To estimate the uncertainty from different
  $b$-fragmentation models, CDF
  compares its \mt\ values using the fragmentation models with the LEP
  parameters~\cite{ALEPH_BFRAG, DELPHI_BFRAG, OPAL_BFRAG} used by D0
  to those using the parameters from the SLD  experiment at SLC~\cite{B-TUNING}. D0
  compares the measured \mt\ with the LEP parameters to the one using parameters from SLD.

  In ATLAS, this uncertainty covers the effects stemming from
  $b$-quark fragmentation, hadronisation and underlying soft
  radiation. It is determined using different Monte Carlo generators
  as well as variations of the $b$-quark fragmentation
  model~\cite{ATLASJES2}.  For the ATLAS \ttbarlj\ input
  measurement~\cite{ATLlj2011}, due to the simultaneous fit of \mt\
  together with JSF and bJSF, the impact of this uncertainty is
  reduced to 0.08~\GeV, albeit at the cost of an additional
  statistical component in the iJES class, which, with the present
  integrated luminosity, amounts to 0.67~\GeV.
  For CMS, the bJES uncertainty on \mt\ is evaluated applying the
  full flavour-dependent JES uncertainty, based on the difference
  in the response between light quark and gluon originated
  jets~\cite{CMSJES}, to $b$-quark originated jets.

  This uncertainty class is assumed to be fully correlated between
  measurements from the same experiments ($\rho_{\rm EXP}=1$). It is
  fully correlated between the
  Tevatron experiments ($\rho_{\rm TEV}=1.0$) and partially correlated
  across LHC experiments ($\rho_{\rm LHC}=0.5$ assumed) because of the different
  methods used to evaluate it. 
  Owing to the methodologies exploited for the estimate of this
  uncertainty source, different correlation assumptions are used as
  the baseline across experiments and colliders: $\rho_{\rm
    ATL-TEV}=1,~\rho_{\rm CMS-TEV}=0.5$. Stability checks are
  performed changing the value of $\rho_{\rm LHC}$ and $\rho_{\rm
    CMS-TEV}=0.5$ to unity (see Section~\ref{sec:comments}).

\end{itemize}

\subsection{Theory and modelling uncertainties}
\label{sec:MCunc}
The component of the
systematic uncertainty stemming from the modelling of \ttbar\ signal events is
divided into several sub-categories. 
Although different baseline MC generators and parameter settings are used
within the four collaborations (Section~\ref{sec:calib} and
Table~\ref{tab:MCsetup}), as default assumption, and unless
otherwise stated, these uncertainty categories are assumed to be fully
correlated among measurements in each experiment ($\rho_{\rm
  exp}=1$), and across experiments and colliders ($\rho_{\rm
  TEV}=\rho_{\rm LHC}=\rho_{\rm COL}=1$). Changes in these
assumptions are considered in the combination stability checks (see
Section~\ref{sec:comments}).

 \begin{itemize}
 \item[MC:] (Monte Carlo) This sub-category includes uncertainties
   stemming from the specific choice of the Monte Carlo generator, and
   the hadronisation and underlying event models. When appropriate, 
   identical sets of hard-scatter events are used in the comparison of
   different MC models.
   
   For CDF, the hadronisation and underlying event (UE) uncertainty is calculated by comparing \mt\
   obtained using \Pythia\ with Tune~A of the underlying event
   model to results from \Herwig\ with a tuned implementation of
   the underlying-event generator Jimmy~\cite{JIMMY, JIMMYtuneCDF}. 
   D0 estimates this uncertainty component by comparing \mt\ results
   using \Alpgen\ interfaced to \Herwig\ relative to the baseline
   \Alpgen+\Pythia, with the corresponding tunes.
   Furthermore, for both CDF and D0 measurements, a contribution to
   this uncertainty category stems from higher order corrections, and
   the simulation of decay widths effects for the top quark and the
   $W$ bosons.  This is obtained by comparing \mt\ results from LO and
   NLO generators, using \Pythia\ (CDF) or \Alpgen\ (D0) versus
   \MCATNLO~\cite{MCNLO, FRI-0301}, and varying the relative fraction of
   $q\bar q \to \ttbar$ and $gg \to \ttbar$ sub-processes in the
   simulation (for CDF, a re-weighting of the gluon fusion fraction in
   the \Pythia\ model from 5\% to 20\% is performed).

   ATLAS estimates the MC generator systematic uncertainty by comparing \mt\
   results obtained with \MCATNLO\ with those from
   \Powheg, both interfaced to \Herwig. In addition, contributions due
   to the choice of the hadronisation model (\Pythia\ versus \Herwig) used in the simulation are
   also included (see also Section~\ref{sec:comments}).
   For CMS, the baseline \Madgraph\ MC setup does not include the
   simulation of the decay widths of the top quarks and the $W$
   bosons. A systematic uncertainty is obtained by comparing the \mt\
   results in MC samples generated with \Powheg\ or \Madgraph\
   to also cover this effect~\cite{LHC2013}.
   Both LHC experiments evaluate separately the UE contribution by
   comparing simulated samples 
   interfaced to \Pythia\ with tunes Perugia 2011 and Perugia 2011
   mpiHi~\cite{MCATLASTUNE}. 
   The UE-specific uncertainty for ATLAS and CMS measurements
   (separated in Ref.~\cite{LHC2013}) is added to this category.

 \item[Rad:] (Radiation) This category includes uncertainties due to
   the modelling of QCD radiation in \ttbar\ events.

   Uncertainties from QCD initial state radiation 
   are assessed by both Tevatron collaborations using a CDF
   measurement~\cite{ISR-FSR-CDF} in Drell-Yan dilepton events that
   have the same $q\bar{q}$ initial state as most \ttbar\ events, but
   no final state radiation. The mean $p_{\rm T}$ of the dilepton
   pairs is measured as a function of the dilepton invariant mass, and
   the values of $\Lambda_{\rm QCD}$ and the $Q^2$ scale in the
   simulation (based on the \Pythia\ parton shower model) that bracket
   the data when extrapolated to the \ttbar\ mass region are
   found. These are used to define the corresponding parameter
   variations.
   The same
   variations of $\Lambda_{\rm QCD}$ and $Q^2$ scale are used to
   estimate the effect of final state radiation.

   For the ATLAS measurements, variations of the initial and final
   state radiation (ISR/FSR) parameters within \Pythia, which are
   constrained by \ttbar-enriched ATLAS data~\cite{LHCISR},
   are used to evaluate these \mt\ systematic uncertainties. In CMS,
   \Madgraph\ MC samples
   with varied factorisation and renormalisation scales, as well as
   with varied $p_{\rm T}$ thresholds for the MLM matching~\cite{MLM},
   are used to address these systematic uncertainties.  Investigations
   from Refs.\cite{LHCISR,ATL_RADPUBNOTE, LHCISR2} indicate that the
   ATLAS and CMS approaches describe, to a large extent, the same
   physics effect.

   Due to the difference between the dominant initial state production
   processes yielding the \ttbar\ pairs at the Tevatron and at the
   LHC, and owing to the different methods applied to
   constrain the radiation related parameters in the MC\footnote{The
     Drell-Yan events used at the Tevatron mainly constrain gluon
     radiation off of (anti)quarks. On the other hand, the $\ttbar$
     dilepton topologies, used in the jet-veto
     analyses~\cite{LHCISR,LHCISR2} at the LHC, mainly constrain QCD
     radiation off of gluons.}, this uncertainty contribution is assumed
   to be partially correlated between measurements from different
   colliders ($\rho_{\rm COL}\equiv \rho_{\rm ATL-TEV} = \rho_{\rm
     CMS-TEV}= 0.5$).  Variations of these assumptions are reported in
   Section~\ref{sec:comments}.

   Some level of double counting between this uncertainty source and
   the stdJES and MC categories described above might be present.

 \item[CR:] (Colour Reconnection) This is the part of the uncertainty
   related to the modelling of colour reconnection effects~\cite{CR}.
 
   This uncertainty is evaluated by comparing \mt\ results obtained in
   simulated samples with the hadronisation based on \Pythia\ tunes APro
   and ACRPro (for CDF and D0)~\cite{TunesAPRO}, or Perugia 2011 and
   Perugia 2011 noCR (for ATLAS and CMS)~\cite{MCATLASTUNE}.
    
 \item[PDF:] (Parton Distribution Functions) This is the part of the
   modelling uncertainty related to the proton PDF. 

   For CDF, the uncertainty is evaluated by comparing
   CTEQ5L~\cite{CTEQ5} results with MRST98L~\cite{MRST98}, by changing
   the value of the strong coupling constant, $\alpha_S$, in the MRST98L model, and by re-weighting
   the simulated events according to the 20 eigenvector variations in
   CTEQ6M~\cite{CTEQ6}. 
   D0 estimates this uncertainty by
   re-weighting the \Pythia\ model to match possible excursions in the
   parameters represented by the 20 CTEQ6M eigenvector variations.

   For ATLAS and CMS, the PDF uncertainty is evaluated by re-weighting
   the simulated signal events according to the ratio of the default
   central PDF (CT10 and CTEQ6.6L for ATLAS and CMS, respectively) and
   the corresponding eigenvector variations~\cite{CT10, CTEQ66, PDF4LHC}.  The
   uncertainty contribution corresponding to the re-weighting of the
   events to alternative PDF sets is found to be smaller than the
   above variation and not included.

   Using the methodology described in Ref.~\cite{HIGGSYELLOWBOOK2},
   the correlation of the effects due to PDF variations on \ttbar\
   production mechanisms between the Tevatron and the LHC is estimated
   to be moderate (about $50-60\%$
   for the LHC run at $\sqrt{s}=7~\TeV$) and to decrease as a function
   of the LHC centre-of-mass energy. While this result could be
   applied for the \mt determinations using \ttbar\ production cross
   section measurements, the correlation of the uncertainties due to
   PDF, across the
   input measurements considered here, might be reduced owing to the
   \ttbar\ event reconstruction. As a result, $\rho_{\rm COL}= 0.5$
   is used as the default correlation
   assumption. Variations of these
   assumptions are reported in Section~\ref{sec:comments}.

\end{itemize}

\subsection{Uncertainties on the detector modelling, background
  contamination, and environment}
\label{sec:otherunc}
The  following systematic uncertainties arise from effects not directly connected 
to the JES and the theoretical modelling of the \ttbar\ signal.

\begin{itemize}

\item[DetMod:] (Detector Modelling) This category relates to
  uncertainties in the modelling of detector effects (uncertainties
  related to the $b$-jet and lepton identification are treated
  separately as detailed below). For Tevatron experiments, this
  systematic uncertainty arises from uncertainties in the modelling of
  the detector in the MC simulation. For D0, this includes the
  uncertainties from jet resolution and jet reconstruction~\cite{JET-ENERGY-SCALE-DZERO-NEW}.  At CDF, the
  jet reconstruction efficiency and resolution in the MC closely match
  those in data~\cite{JET-ENERGY-SCALE-CDF}. 
  The small differences
  propagated to \mt, after increasing the jet resolution in the MC by
  $\approx 4\%$ in absolute value, lead to a negligible uncertainty ($<$0.01 \GeV).
  For LHC experiments, this category includes uncertainties in the
  jet energy resolution~\cite{ATLASJETRES, CMSJES}, the jet
  reconstruction efficiency~\cite{ATLASJESPAPER2010} as well as uncertainties
  related to the reconstruction of the  missing transverse energy,
  \met~\cite{ATLASMET, CMSMET}.  This uncertainty class is
  assumed to be fully correlated between measurements from the same
  experiments ($\rho_{\rm EXP}=1$), but uncorrelated across experiments
  ($\rho_{\rm TEV}=\rho_{\rm LHC}=\rho_{\rm COL}=0$).

\item[$b$-tag:] ($b$-tagging) This is the part of the uncertainty
  related to the modelling of the $b$-tagging efficiency and the
  light-quark jet rejection factors in the MC simulation with respect
  to the
  data~\cite{CDFBTAG,CDF-D0-Combo2012,ATLASBTAG,ATLASBTAG2,ATLASBTAG3,CMSBTAG,D0BTAG}.

  CDF reports that any difference between the $b$-tagging behaviour in
  MC and data~\cite{CDFBTAG} has a marginal impact on the
  measurement of \mt: the latest CDF measurements in the \ttbarlj\ and 
  \ttbaraj\ channels estimated this effect to be 0.03~\GeV\ and
  0.1~\GeV, respectively. It is not evaluated for the \ttbarll\ and \ttbarmet\
  results (``n.e.'' in Table~\ref{tab:syscat}).
  D0 also exploits $b$-tagging algorithms in the analyses~\cite{D0BTAG}:
  the $b$-tagging
  efficiency for simulated events is adjusted to match the data, and
  the corresponding uncertainties are propagated to the \mt\ analyses.
  The \mt\ combined result is obtained with $\rho_{\rm EXP} =1$ and
  $\rho_{\rm TEV}=0$.  

  For LHC experiments, data-to-MC $b$-tagging scale factors (SF) are
  derived as a function of the jet properties (flavour, $p_{\rm T}$,
  and $\eta$) using $b/c/\rm light$-quark enriched dijet data
  samples. In some cases (ATLAS \ljets\ analysis) SF are derived from
  a combination of different calibrations obtained from a \ttbarll
  sample~\cite{ATLASBTAG2}, and a sample of jets
  including muons~\cite{ATLASBTAG3}; nevertheless, as
  default assumption, we used $\rho_{\rm ATL} = \rho_{\rm CMS} =1$ in
  the present results.
  As opposed  to what was done in
  Ref.~\cite{LHC2013}, the present combination is carried out with $\rho_{\rm
    LHC}=0$ for this source.  Variations of this assumption are
  analysed in the stability checks (see Section~\ref{sec:comments} for
  further details). Finally $\rho_{\rm COL}=0$ is used for the default result.
  Despite the sizable reduction of the bJES related systematics that
  is achieved, the ATLAS \ttbarlj\ analysis exhibits an increased
  sensitivity to the uncertainties of the $b$-tagging efficiency and
  of the light jet rejection factors. This is related to the shape
  differences introduced by the $b$-tagging SF variations in
  the variable sensitive to the bJSF~\cite{ATLlj2011}.

\item[LepPt:] This category takes into account the uncertainties in
  the efficiency of the trigger, in the identification and
  reconstruction of electrons and muons, and residual uncertainties
  due to a possible miscalibration of the lepton energy and momentum
  scales~\cite{CDF-D0-Combo2012,ATLASel,ATLASmu,CMSmu}. The
  correlation assumptions for this uncertainty source are
  $\rho_{\rm EXP}=1$, and $\rho_{\rm TEV}=\rho_{\rm LHC}=\rho_{\rm COL}=0$.

\item[BGMC:] (Background from MC) This represents the uncertainty due
  to the modelling of the background processes 
  determined from MC. This uncertainty source is assumed to be fully
  correlated between all input measurements in the
  same \ttbar\ decay channel
  ($\rho_{\rm EXP}=\rho_{\rm TEV}=\rho_{\rm LHC}=\rho_{\rm COL}=1$ in
  the same analysis channel).

\item[BGData:] (Background from data) This class includes the
  uncertainties of the modelling of the background determined from
  data, and is assumed to be uncorrelated between all input
  measurements ($\rho_{\rm EXP}=\rho_{\rm TEV}=\rho_{\rm
    LHC}=\rho_{\rm LHC} = 0$).  These typically originate from
  uncertainties in the normalisation of the QCD multijet and Drell-Yan
  backgrounds determined from data.

\item[Meth:] (Method) This systematic uncertainty relates to the \mt\
  extraction technique adopted by the analyses (uncorrelated between
  all measurements: $\rho_{\rm EXP}=\rho_{\rm TEV}=\rho_{\rm
    LHC}=\rho_{\rm COL}=0$). This includes uncertainties caused by the
  limited MC statistics available for the measurement calibration.

\item[MHI:] (Multiple Hadronic Interactions) This systematic
  uncertainty arises from the modelling of the pile-up conditions in
  the simulation with respect to the data (overlay of multiple
  proton-(anti)proton interactions).  It is assumed to be fully correlated
  between all measurements in the same experiments
  ($\rho_{\rm EXP}=1$). Between the Tevatron experiments this uncertainty
  is treated as uncorrelated due to the different methods of
  determination ($\rho_{\rm TEV}=0$)\footnote{D0 models the pile-up
    contribution overlaying real minimum-bias data events in the MC simulation.}. For LHC experiments, this
  uncertainty is MC driven and assumed to be
  fully correlated ($\rho_{\rm LHC}=1$). Finally, for this category
  $\rho_{\rm COL} = 0$ is used in the combination.

\end{itemize}

\newpage

\section{World \mt\ combination}
\label{sec:results}

Using the BLUE method, and the information described above, the
combined value of \mt\ is: 

\noindent

$$ \mt = 173.34 \pm 0.27 \mbox{ (stat)} \pm 0.71 \mbox{
  (syst)}~\GeV.$$

\noindent Alternatively, separating the iJES statistical contribution
from the quoted systematic uncertainty, the result reads:
  
$$ \mt = 173.34 \pm 0.27 \mbox{ (stat)} \pm 0.24 \mbox{ (iJES)} \pm 0.67 \mbox{ (syst)}~\GeV.$$

\noindent
The $\chi^2$ of the combination is 4.3 for 10 degrees of freedom (ndf)
and the corresponding probability is 93\%~\cite{BLUE1, BLUE2}. This
value, calculated taking correlations into account, can be used to
assess the extent to which the individual measurements are consistent
with the combined \mt\ value and with the hypothesis that they measure
the same physics parameter.  Moreover, for each input value $m_i$
(with an overall uncertainty $\sigma_i$), the
pull, calculated as ${\rm pull}_i = (m_i - \mt)/\sqrt{\sigma_i^2 -
  \sigma_{\mt}^2}$, where $\sigma_{\mt}$ is the total uncertainty of
the combined \mt\ result, indicates the degree of agreement among the
input measurements.

\begin{figure}
\begin{center}
\subfigure[]{\label{fig:summaryA} \includegraphics[width=0.78\textwidth]{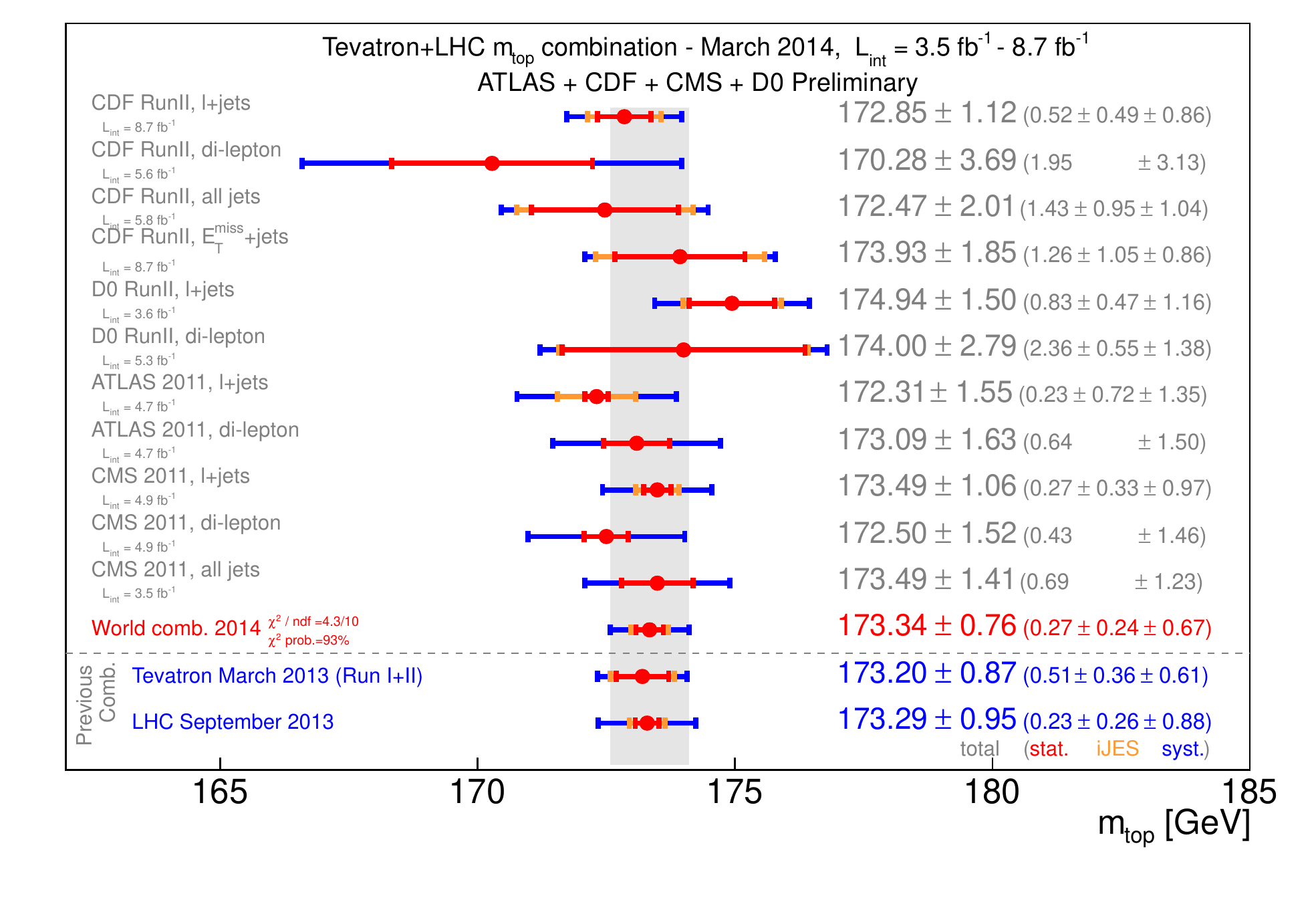}}
\subfigure[]{\label{fig:summaryB} \includegraphics[width=0.30\textwidth]{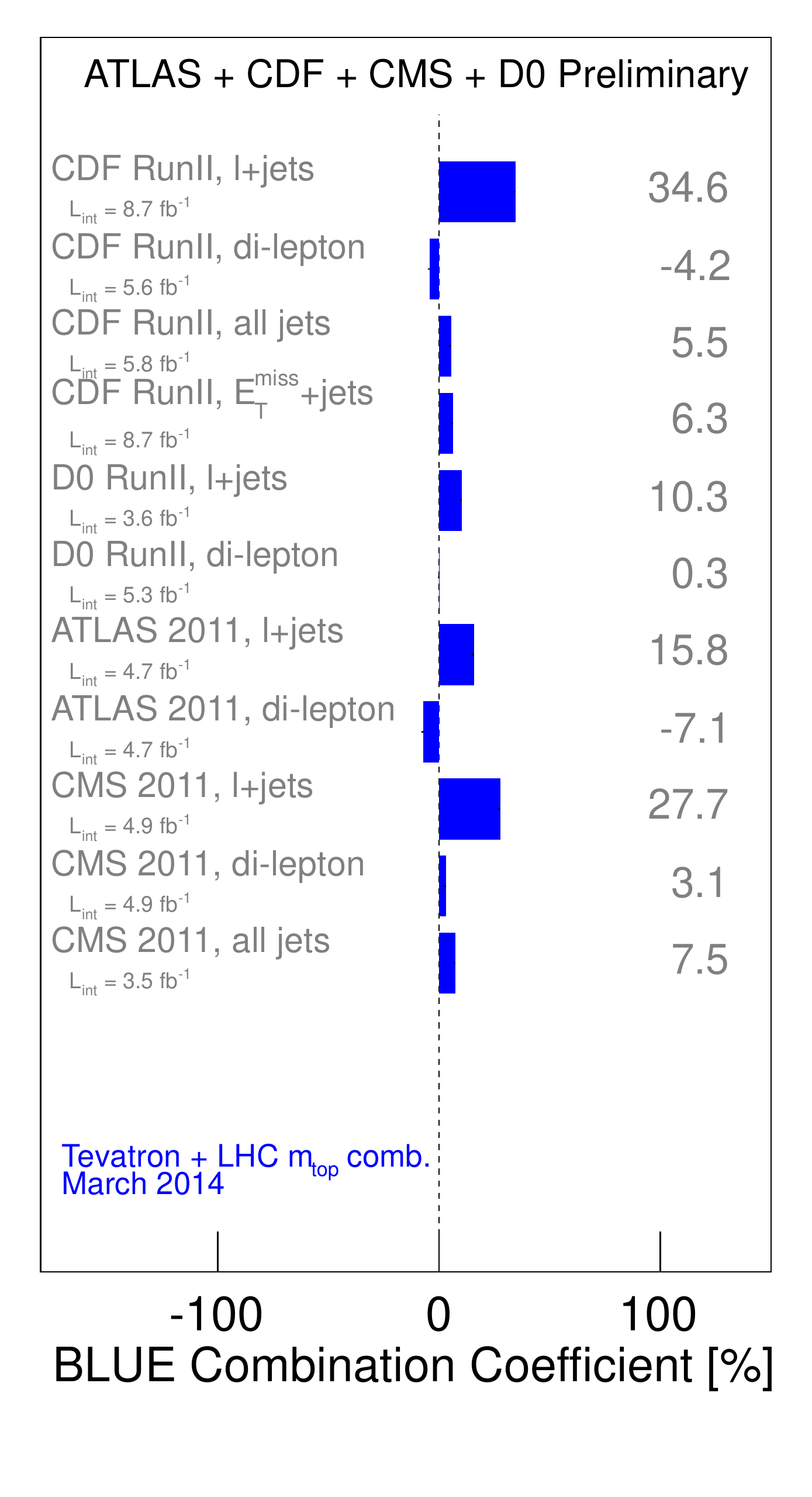}}
\hspace{1.95cm}
\subfigure[]{\label{fig:summaryC} \includegraphics[width=0.30\textwidth]{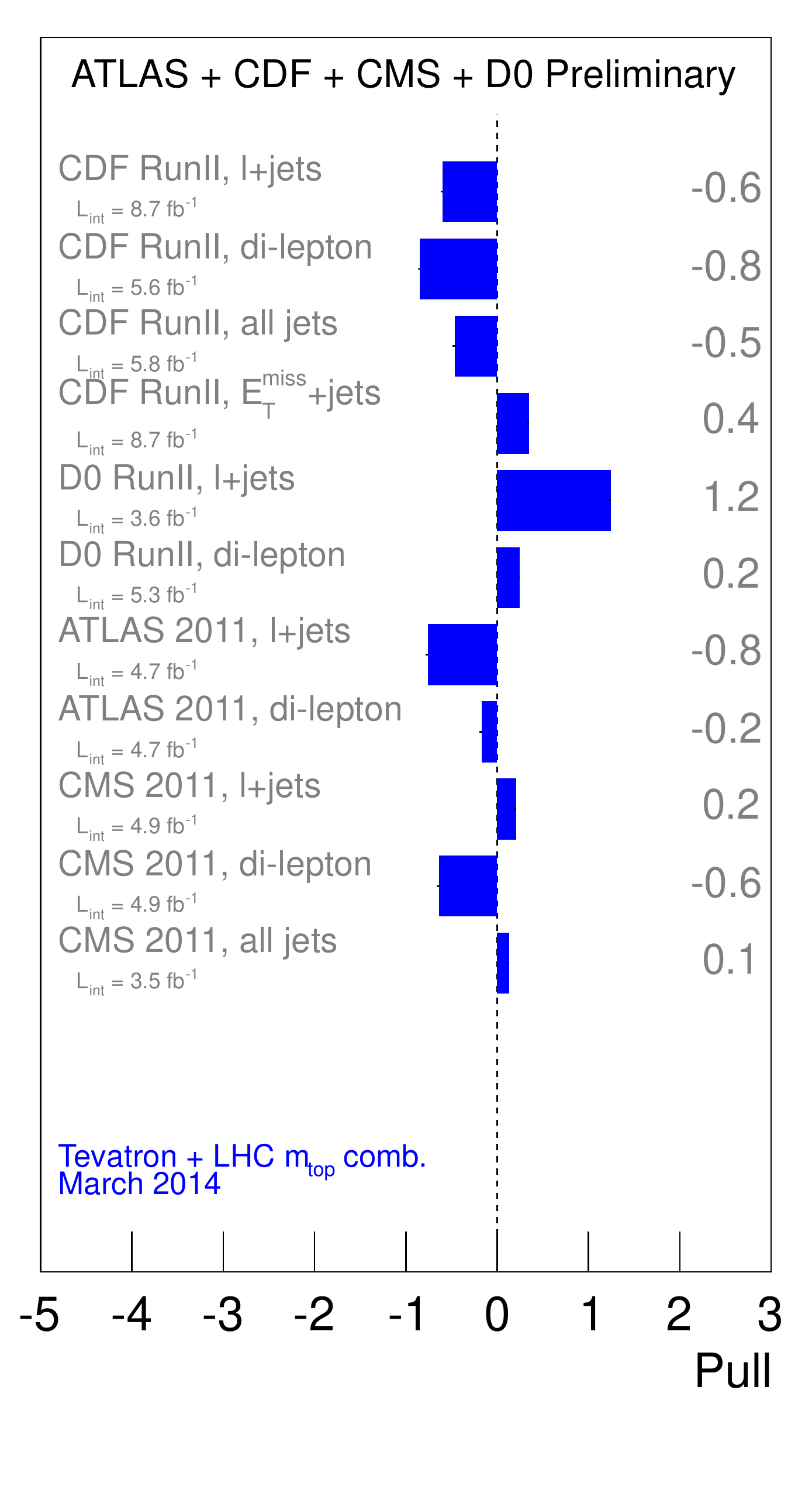}}

\end{center}
\caption{(a): Input measurements and result of their combination (see
  also Table~\ref{tab:syscat}), compared with the Tevatron and LHC
  combined \mt\ values~\cite{TEV2013,LHC2013}. With respect to
  Ref.~\cite{TEV2013} only a partial set of Tevatron \mt\ measurements
  is used in the world combination (see Section~\ref{sec:input}).  For
  each measurement, the total uncertainty, the statistical and the
  iJES contributions (when applicable), as well as the sum of the
  remaining uncertainties are reported separately. The iJES
  contribution is statistical in nature and applies only to analyses
  performing in situ (\ttbar) jet energy calibration procedures.  The
  grey vertical band reflect the total uncertainty on the combined
  \mt\ value. Panels (b)
  and (c) show, respectively, the BLUE combination coefficients and pulls of the input
  measurements. }
\label{fig:summary}
\end{figure}
Table~\ref{tab:syscat} and Figure~\ref{fig:summary} summarise the
inputs and the results of the combination.
In Figure~\ref{fig:summaryA}, this \mt\ combination result, the individual
Tevatron and LHC combinations~\cite{TEV2013, LHC2013} and the input
measurements are compared.
The total uncertainty, the statistical and the iJES contributions (when applicable),
as well as the sum of the remaining uncertainties are reported separately.
The central value of the world \mt\ combination falls outside the range defined by the
central values of the individual Tevatron and LHC combinations from Refs.~\cite{TEV2013,
  LHC2013}. This is a consequence of the reduced set of input
measurements used in this combination with respect to
Ref.~\cite{TEV2013} (see below for further details).
It may be worth noting that the statistical uncertainty (including in
quadrature the iJES component) of the world \mt\ combination is
slightly larger than the corresponding uncertainty reported in the
previous LHC combination~\cite{LHC2013}. A similar consideration holds
for the statistical uncertainty of the input measurements relative to
that of the combined result. This can happen due to the combination
method, which minimises the total uncertainty of the combined
result, not the separate statistical and systematic contributions
(see Section~\ref{sec:blue}). In
this framework, the breakdown of uncertainties for the combined
result, is a function not only of the uncertainties of the input
measurements, but also of the correlations among them through the
combination~\cite{BLUE2}.
The BLUE combination coefficients used in the linear combination of
the input \mt\ values and the pulls are provided in
Figures~\ref{fig:summaryB} and \ref{fig:summaryC}, respectively.
Within the BLUE method, negative coefficients can occur when individual
measurements have different precisions and large
correlations~\cite{BLUE1, BLUEFIN, BLUERN}.

The correlations among input $\mt$ measurements are reported in
Table~\ref{tab:inputcorr}.  
The precision of the combined result relative to the most precise
single measurement is improved by about 28\%. The total
uncertainty on \mt\ is 0.76~\GeV, and corresponds to
a relative uncertainty of ~0.44\%.
The resulting total uncertainty is dominated by systematic
contributions related to the modelling of the \ttbar\ signal and the
knowledge of the jet energy scale for light- and $b$-quark originated
jets (Table~\ref{tab:syscat}).

\begin{table}[!t]
\begin{center}
\footnotesize 
\begin{tabular}{|rl||cccc|cc|cc|ccc|}
\hline
&&\multicolumn{4}{|c|}{\multirow{2}[1]{*}{CDF}} &\multicolumn{2}{|c|}{\multirow{2}[1]{*}{D0}}&\multicolumn{2}{|c|}{\multirow{2}[1]{*}{ATLAS}} &\multicolumn{3}{|c|}{\multirow{2}[1]{*}{CMS}} \\
&&&&&&&&&&&&\\
&& \rotatebox{90}{\ljets} & \rotatebox{90}{\dil}&
\rotatebox{90}{\alljets} & \rotatebox{90}{\met} &
\rotatebox{90}{\ljets} & \rotatebox{90}{\dil}&
\rotatebox{90}{\ljets} & \rotatebox{90}{\dil}&
\rotatebox{90}{\ljets} & \rotatebox{90}{\dil} & \rotatebox{90}{\alljets} \\
\hline\hline
 \multirow{4}[1]{*}{CDF} &\ljets          & 1.00 &       &       &       &       &       &       &       &       &       &       \\
   &\dil            & 0.49 & 1.00 &       &       &       &       &       &       &       &       &       \\
   &\alljets        & 0.28 & 0.25 & 1.00 &       &       &       &       &       &       &       &       \\
   &\met            & 0.31 & 0.27 & 0.17 & 1.00 &       &       &       &       &       &       &       \\
\hline                
  \multirow{2}[1]{*}{D0} &\ljets           & 0.29 & 0.09 & 0.16 & 0.18 & 1.00 &       &       &       &       &       &       \\
   &\dil             & 0.15 & 0.07 & 0.10 & 0.11 & 0.38 & 1.00 &       &       &       &       &       \\
\hline                
  \multirow{2}[1]{*}{ATLAS} &\ljets          & 0.17 & 0.07 & 0.10 & 0.12 & 0.17 & 0.11 & 1.00 &       &       &       &       \\
  & \dil            & 0.30 & 0.12 & 0.17 & 0.19 & 0.24 & 0.15 & 0.64 & 1.00 &       &       &       \\
\hline                
  \multirow{3}[1]{*}{CMS} &\ljets          & 0.23 & 0.12 & 0.15 & 0.16 & 0.21 & 0.16 & 0.24 & 0.34 & 1.00 &       &       \\
  & \dil            & 0.09 & 0.05 & 0.05 & 0.08 & 0.08 & 0.07 & 0.16 & 0.24 & 0.64 & 1.00 &       \\
  & \alljets         & 0.15 & 0.06 & 0.09 & 0.10 & 0.13 & 0.08 & 0.15 & 0.23 & 0.57 & 0.75 & 1.00 \\ 
\hline

\end{tabular}
\end{center}

\caption{Correlations among the eleven input measurements.   The elements in the
table are labelled according to the experiment and the \ttbar\ final state.}
\label{tab:inputcorr}

\footnotesize
\begin{center}
\begin{tabular}{|rl|c|c|c|}
\hline
\multicolumn{2}{|c|}{\multirow{2}[1]{*}{Measurements}}& BLUE comb.  &
IIW   & MIW \\
&  & coeff. [\%] & [\%] & [\%] \\
\hline

 \multirow{4}[1]{*}{CDF}  & \ljets          &  34.6  &  46.6 &   16.1 \\
                          & \dil            &   $-$4.2  &   4.3 &    3.0 \\
                          &\alljets        &   5.5  &  14.4 &    1.9 \\
                          &\met            &  6.3    & 16.9 &    2.1  \\  \hline                                               
 \multirow{2}[1]{*}{D0}   & \ljets           &  10.3  &  25.8 &    3.2 \\
                          &\dil            &   0.3  &   7.5 &    0.0 \\ \hline                                              
 \multirow{2}[1]{*}{ATLAS}& \ljets        &  15.8  &  24.2 &    6.1 \\
                          & \dil          &   $-$7.1  &  21.9 &    1.2 \\  \hline                                              
 \multirow{3}[1]{*}{CMS}  &  \ljets          &  27.7  &  51.3 &    7.6  \\
                          & \dil            &   3.1  &  25.1 &    0.1 \\
                          &\alljets        &   7.5  &  29.2 &    0.8 \\
  \hline
\multicolumn{2}{|c|}{Correlations (IIW$_{\rm corr}$)} &  --- &    $-$167.3 & --- \\

\hline
\end{tabular}

\caption{Evaluation of the impact of the individual measurements on
  the combined \mt. The values of the BLUE
 combination coefficients, the intrinsic information weights IIW$_i$,
 and the marginal information weights MIW$_i$ are given. The intrinsic information
 weight IIW$_{\rm corr}$ of correlations is also shown on a
 separate row~\cite{BLUEFIN}. 
 }
\label{tab:BLUEFINweights}
\end{center}
\end{table}

Complementing the information encoded in the BLUE combination
coefficients, the impact of the various input measurements is
estimated using the Fisher information concept, $I =
1/\sigma_{\mt}^2$~\cite{BLUEFIN}. 
For each of the input
measurements, intrinsic (IIW$_i$) and marginal information weights
(MIW$_i$) are derived. The intrinsic information weight carried by the
$i^{\rm th}$-measurement is supplemented by the introduction of a
weight inherent to the ensemble of all correlations between the input
measurements (IIW$_{\rm corr}$):

$${\rm IIW}_i = \frac{1/\sigma_i^2}{1/\sigma_{\mt}^2} =
\frac{1/\sigma_i^2}{I} ; ~~ {\rm IIW}_{\rm corr} = \frac{I - \sum_i 1/\sigma_i^2}{I}. $$

\noindent
The IIW$_i$ for each individual measurement is defined as the ratio of the
information it carries when taken alone ($1/\sigma_i^2$)
to the total
information of the combination. While the IIW$_i$ are defined to be
positive, IIW$_{\rm corr}$ can be negative, or positive, depending on
whether the net effect of the correlations is to increase, or decrease, the total
uncertainty of the combination. 
The marginal information weight, defined as

$${\rm MIW}_i = \frac{I_{\rm~ {\it n}~meas}  - I_{\rm {\it n}-1~meas.:~all~but~
    \it{i}}}{I_{\rm {\it n}~meas}}$$

\noindent
can also be used to quantify the information that an individual
measurement brings in a combination. 
MIW$_i$ quantifies the additional
information brought by the $i^{\rm th}$-measurement when added to a combination
that includes the other $n-1$ inputs.

The intrinsic and marginal information weights, for each individual
input measurement, and the intrinsic information weight of the
correlations, are listed in Table~\ref{tab:BLUEFINweights}. For
comparison, the corresponding BLUE combination coefficients are also
reported.
The intrinsic information weight carried by the ensemble of the
correlations among measurements, IIW$_{\rm corr}$, is large in comparison
to the contribution of the individual \mt inputs (IIW$_i$).  It is
therefore important to monitor the stability of the result under
variations of the correlation assumptions (see
Section~\ref{sec:comments}).
While the exact ranking of the input \mt\ measurements varies
depending on the figure of merit adopted (BLUE combination
coefficient, ${\rm IIW}_i$, or ${\rm MIW}_i$), Table~\ref{tab:BLUEFINweights}
shows that the current combination result is mainly driven by the \mt\
results in the \ttbarlj\ decay channel.

\begin{table}[!t]
\begin{center}
\scriptsize
\begin{tabular}{|l|c|c|c|c|c|c||c|c|c|c|}
\hline

                &Individual    & Parameter           & \multicolumn{4}{|c||}{Correlations} & \multicolumn{4}{c|}{$\chi^2$/ndf ($\chi^2$ probability)}\\
                &comb. [GeV]   &   value [GeV]             &
                $m^{\ljets}$ & $m^{\dil}$ & $m^{\alljets}$ &
                $m^{\met}$ & $m^{\ljets}$ & $m^{\dil}$ &
                $m^{\alljets}$ & $m^{\met}$  \\ \hline
$m^{\ljets}$    & $173.29 \pm 0.80$ & $173.23 \pm 0.78$   & 1.00 &      &      &     & $-$           &      &      &\\ 
$m^{\dil}$      & $172.74 \pm 1.15$ & $172.73 \pm 1.09$   & 0.71 & 1.00 &      &     & 0.43/1 (0.51) & $-$  &      &\\
$m^{\alljets}$  & $173.17 \pm 1.20$ & $173.35 \pm 1.13$   & 0.58 & 0.66 & 1.00 &     & 0.02/1 (0.90) & 0.46/1 (0.50) & $-$ &\\ 
$m^{\met}$ & $173.93 \pm 1.85$ & $174.03 \pm 1.80$   & 0.29 & 0.26 & 0.22 & 1.00& 0.21/1 (0.65) & 0.49/1 (0.48) & 0.13/1 (0.72) & $-$ \\
\hline
\end{tabular}
\end{center}
\caption{Individual and correlated combination results according to
  the various  \ttbar\ final states. The
  correlated determination of the \mt\ per decay channel (parameter
  value) is reported together
  with the pair-wise correlation coefficients, and the compatibility
  tests in terms of $\chi^2$/ndf and its associated probability. For comparison, the results of the separate
  combinations (individual comb.) of the individual inputs from
  Table~\ref{tab:syscat} are reported in the second column.}
\label{tab:channels}

\end{table}

\begin{table}[!t]
\begin{center}
\scriptsize 
\begin{tabular}{|l|c|c|c|c|c|c||c|c|c|c|}
\hline

           &Individual     & Parameter           & \multicolumn{4}{c||}{Correlations} & \multicolumn{4}{c|}{$\chi^2$/ndf ($\chi^2$ probability)}\\
           & comb. [GeV]   &   value [GeV]            &  $m^{\rm CDF}$
           & $m^{\rm D0}$ & $m^{\rm ATL}$ &  $m^{\rm CMS}$ & $m^{\rm
             CDF}$ & $m^{\rm D0}$ & $m^{\rm ATL}$ & $m^{\rm CMS}$  \\ \hline
$m^{\rm CDF}$  & $ 173.19\pm 1.00$ & $172.96 \pm 0.98$  & 1.00 &      &      &     & $-$           &      &      &\\ 
$m^{\rm D0}$   & $ 174.85\pm 1.48$ & $174.62 \pm 1.46$  & 0.31 & 1.00 &      &     & 1.25/1 (0.27) & $-$  &      &\\
$m^{\rm ATL}$  & $ 172.65\pm 1.44$ & $172.70 \pm 1.43$  & 0.29 & 0.23 & 1.00 &     & 0.03/1 (0.86) & 1.14/1 (0.29) & $-$ &\\ 
$m^{\rm CMS}$  & $ 173.58\pm 1.03$ & $173.54 \pm 1.02$  & 0.25 & 0.22 & 0.32 & 1.00& 0.23/1 (0.64) & 0.46/1 (0.50) & 0.32/1 (0.57) & $-$ \\
\hline
\end{tabular}
\end{center}

\caption{Individual and correlated combination results according to
  the various  experiments. The
  correlated determination of the \mt\ per experiment (parameter
  value) is reported together
  with the pair-wise correlation coefficients, and the compatibility
  tests in terms of $\chi^2$/ndf and its associated probability. For comparison, the results of the separate
  combinations (individual comb.) of the individual inputs from
  Table~\ref{tab:syscat} are reported in the second column.}
\label{tab:exp}

\end{table}

\begin{table}[!t]

\begin{center}
\scriptsize 
\begin{tabular}{|l|c|c|c|c||c|c|}
\hline

           &Individual     & Parameter           & \multicolumn{2}{c||}{Correlations} & \multicolumn{2}{|c|}{$\chi^2$/ndf ($\chi^2$ probability)}\\
           & comb. [GeV]   &   value [GeV]       &  $m^{\rm TEV}$ &
           $m^{\rm LHC}$  &  $m^{\rm TEV}$ & $m^{\rm LHC}$ \\ \hline
$m^{\rm TEV}$  & $173.58 \pm 0.94 $ & $173.41 \pm 0.91$  & 1.00 &      &  $-$           &     \\ 
$m^{\rm LHC}$  & $173.28 \pm 0.94 $ & $173.26 \pm 0.94$  & 0.36 & 1.00 &  0.02/1 (0.89) & $-$ \\

\hline
\end{tabular}
\end{center}

\caption{Individual and correlated combination results according to
  the Tevatron and LHC colliders. The
  correlated determination of the \mt\ per collider (parameter
  value) is reported together
  with the pair-wise correlation coefficients, and the compatibility
  tests in terms of $\chi^2$/ndf and its associated probability. For comparison, the results of the separate
  combinations (individual comb.) of the individual inputs from
  Table~\ref{tab:syscat} are reported in the second column. 
  See text for further details.}

\label{tab:col}

\end{table}

\begin{figure}[h!]
\begin{center}
\includegraphics[width=0.78\textwidth]{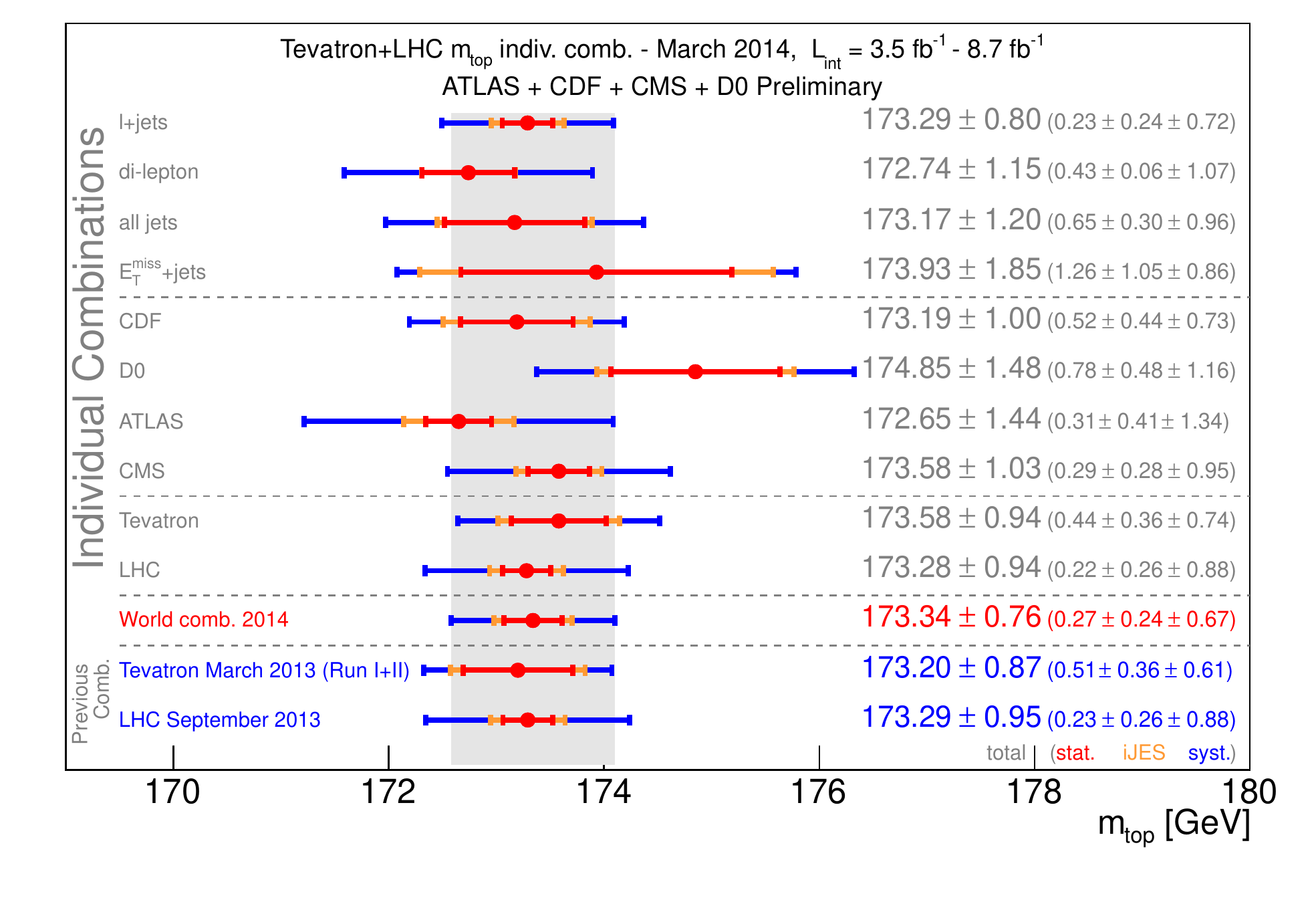}

\end{center}
\caption{Comparison of the world \mt\ combination result with the
  individual \mt\ determinations per \ttbar\ decay channel, experiment,
  and collider. Results are compared with the Tevatron and LHC
  combined \mt\ values from Refs.~\cite{TEV2013,LHC2013}. The
  grey vertical band reflect the total uncertainty on the combined
  \mt\ value. }
\label{fig:sepcomb}
\end{figure}

Using the same inputs, uncertainty categorisation, and correlation
assumptions, additional combinations have been performed as detailed
below.

\begin{itemize}
\item {\bf Individual combinations}  by \ttbar\ final
  state, experiment, and collider, have been derived neglecting
  other input measurements. These results can be used to quantify
  the improvement obtained by the overall combination with respect to
  the results obtained using only a partial set of the input measurements.

\item{\bf Correlated combinations} are obtained within the BLUE
  program by simultaneously extracting different mass parameters
  instead of a common \mt. The \mt\ parameter values obtained using
  this procedure (per \ttbar\ final state, experiment, or collider),
  are affected by the full set of input measurements through their
  correlations, and can be used to test the consistency between the
  various \mt\ determination. In the following this is done using a pair-wise
  $\chi^2$ formulation and its associated probability: $\chi^2(m_1,
  m_2) = (m_1-m_2)^2/ \sigma_{12}^2$, where $\sigma_{12}^2=
  \sigma_{1}^2 + \sigma_{2}^2 - 2\rho_{12}\sigma_{1}\sigma_{2}$, and
  $\rho_{12}$ is the correlation between the two measurements.
\end{itemize}

The results are summarised in Tables~\ref{tab:channels},
\ref{tab:exp} and \ref{tab:col}, respectively, for the combination
according to the \ttbar\ final state ($m^{\ljets}$, $m^{\dil}$, $m^{\alljets}$,
$m^{\met}$), to the individual experiments ($m^{\rm CDF}$,
$m^{\rm D0}$, $m^{\rm ATL}$, $m^{\rm CMS}$), and to the Tevatron and LHC
colliders ($m^{\rm TEV}$, $m^{\rm LHC}$).
Figure~\ref{fig:sepcomb} reports the comparison of the world \mt\
combination with the individual \mt\ determinations per \ttbar\ decay
channel, experiment, and collider. In addition \mt\ combination
results from Refs.~\cite{TEV2013,LHC2013} are also reported (see
Appendix~\ref{app:extrafigs}, Figure~\ref{fig:corrcomb} for the correlated
\mt\ determinations).
The full uncertainty breakdown of the individual CDF, D0, ATLAS, CMS,
Tevatron and LHC combinations is reported in
Appendix~\ref{app:sepcomb}.
The individual combination for $m^{\rm TEV}$ and $m^{\rm LHC}$ present
some differences with respect to the results documented in
Refs.~\cite{TEV2013, LHC2013}. For $m^{\rm TEV}$, these mainly
originate from the reduced set of input measurements used in the
combination with respect to Ref.~\cite{TEV2013}, and to a lesser
extent from the use of a finer MC modelling uncertainty splitting
(four separate categories: MC, Rad, CR, PDF, rather than a single one
including all of them), and the change in the JES uncertainty
categories for the CDF measurements. The slight differences in the
uncertainty breakdown of the separate combination of $m^{\rm LHC}$
with respect to Ref.~\cite{LHC2013} are mainly attributed to the
changes of the uncertainty categorisation and correlation assumption
underlying the stdJES and $b$-tagging categories.

\section{Effects of using alternative correlation models and uncertainty treatments}
\label{sec:comments}

\begin{figure}
\begin{center}
\subfigure[]{\label{fig:checkA}\includegraphics[width=0.48\textwidth]{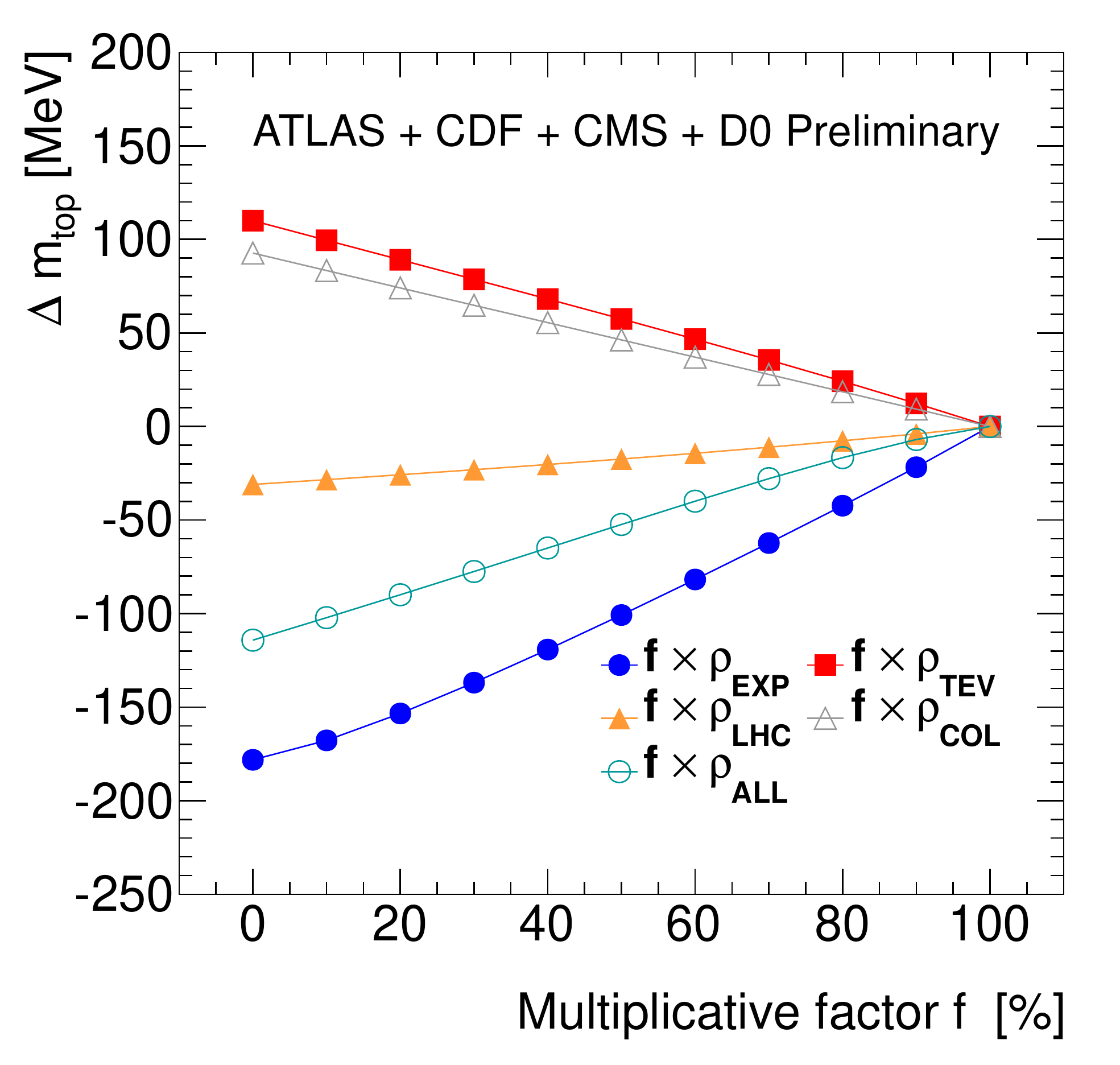}}
\subfigure[]{\label{fig:checkB}\includegraphics[width=0.48\textwidth]{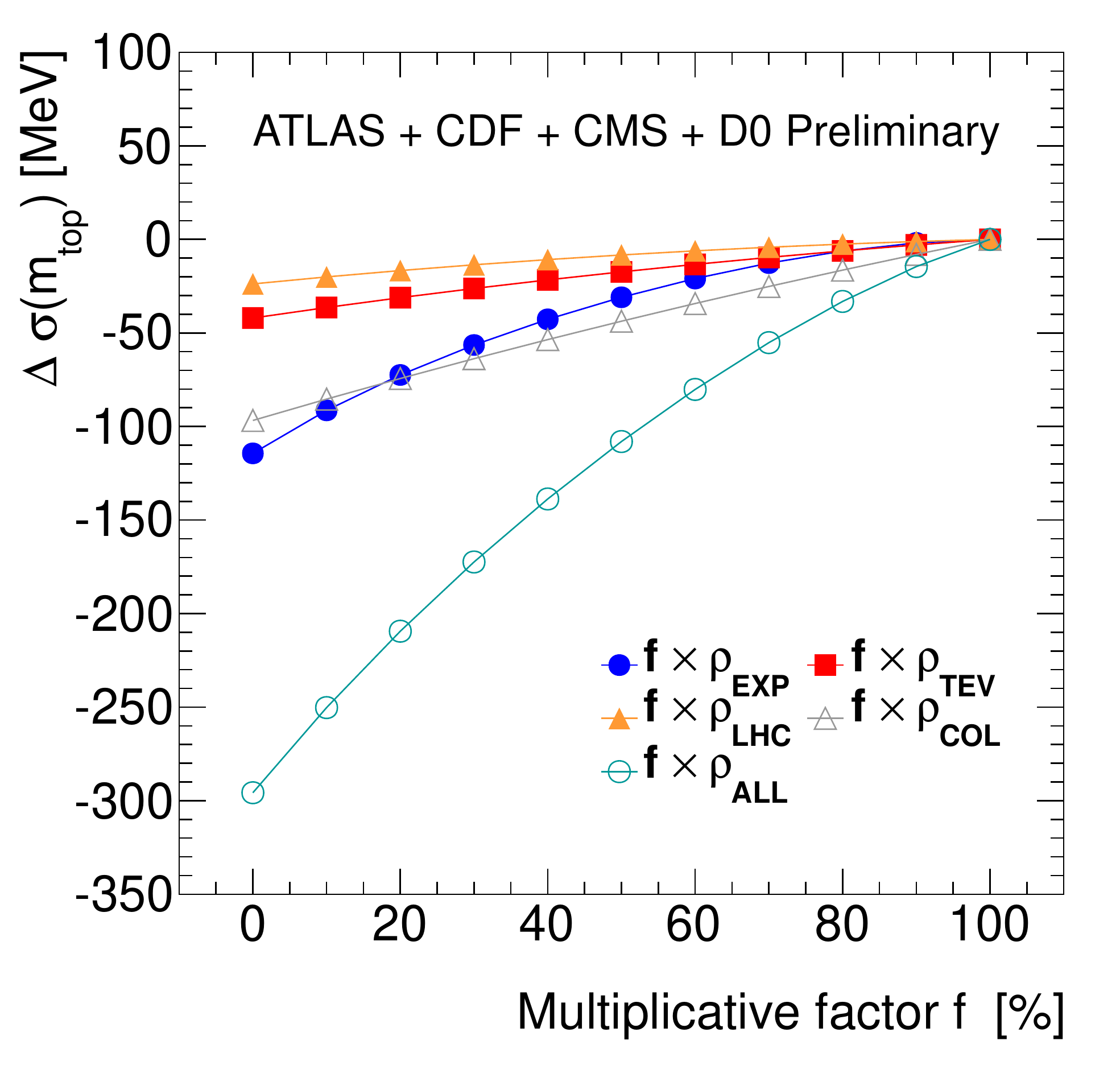}}
\hspace{1.1cm}
\subfigure[]{\label{fig:checkC}\includegraphics[width=0.48\textwidth]{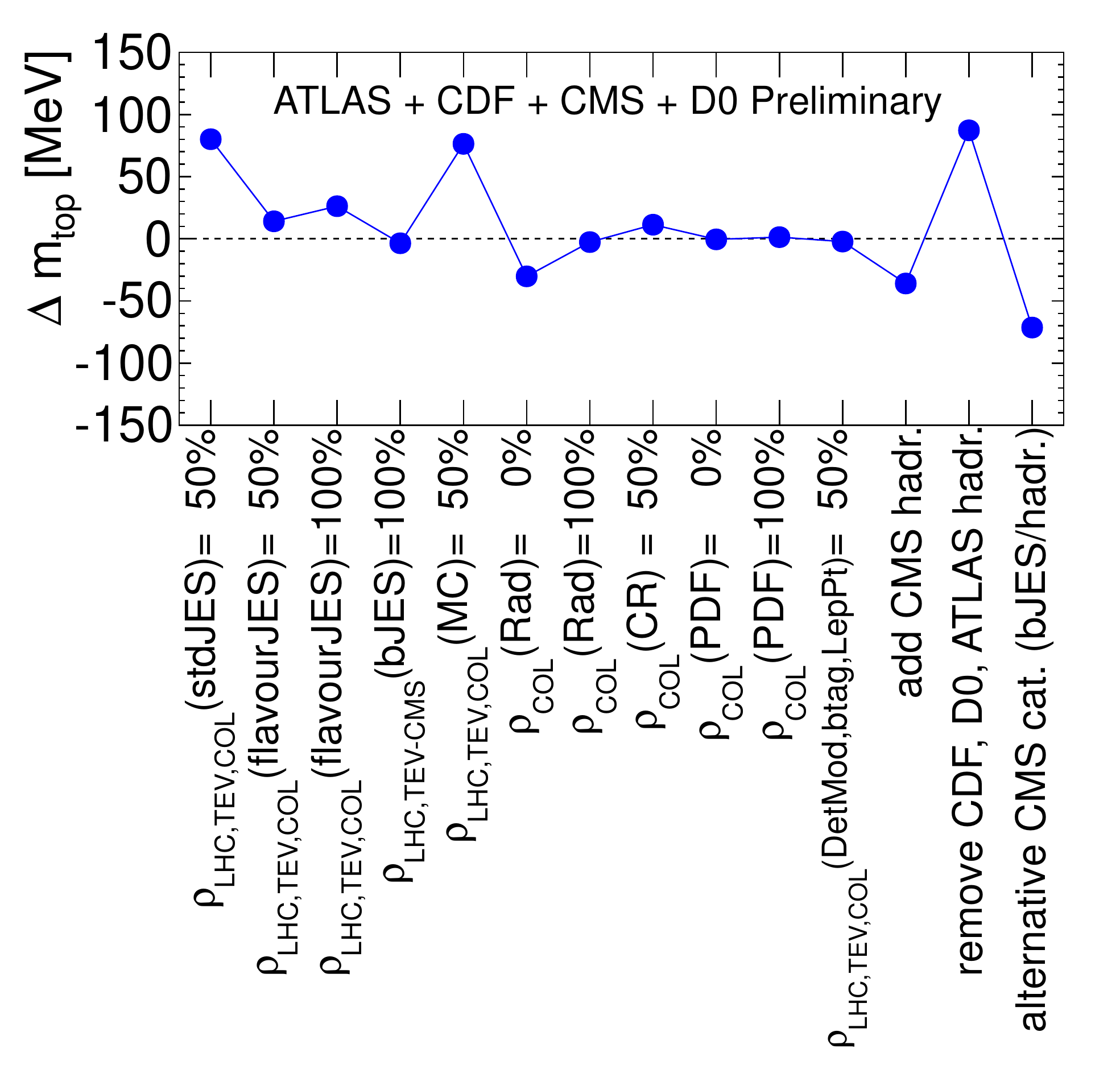}}
\subfigure[]{\label{fig:checkD}\includegraphics[width=0.48\textwidth]{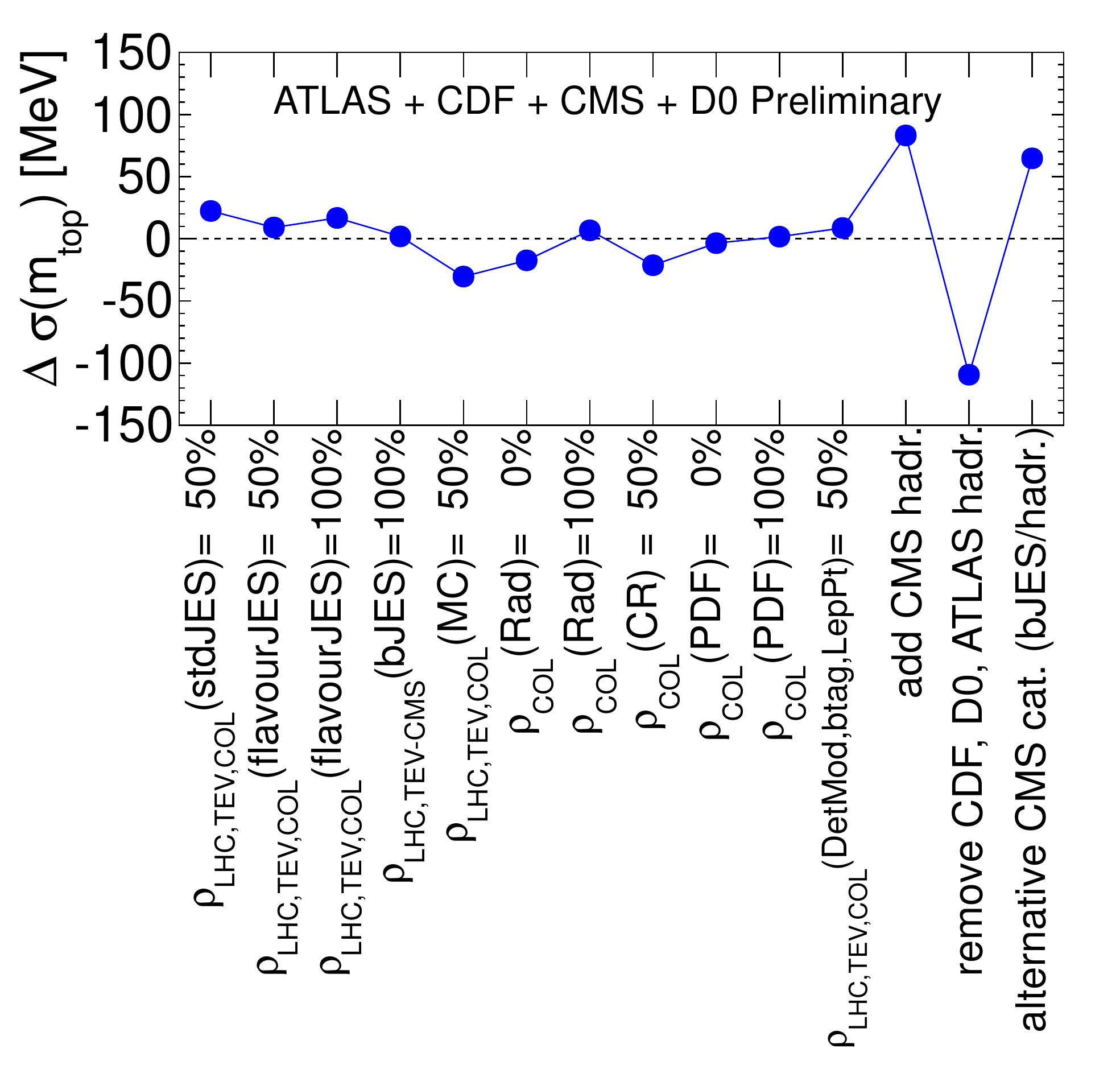}}

\end{center}
\caption{Variation of the combined \mt result (a,c) and its total
  uncertainty (b,d) as a function of variations in the correlation
  assumptions.  (a,b) $\rho_{\rm EXP}$, $\rho_{\rm LHC}$, $\rho_{\rm
    TEV}$ and $\rho_{\rm COL}$ are varied simultaneously using a
  multiplicative factor $f$ in the range [0,1] (open light blue
  dots). Separate variations of each correlation coefficient in the
  same range, are reported by the blue (filled dots), orange (filled
  triangles), red (filled squares) and the grey (open triangles)
  curve, respectively.  (c,d) Stability of the world combination under
  variations of the default assumptions on the correlation for
  selected uncertainty sources. The sensitivity of the combination to
  different scenarios concerning the treatment of the hadronisation
  systematics is also shown. See text for details.}
\label{fig:checks}
\end{figure}

The categorisation and the correlation assumptions summarised in
Tables~\ref{tab:syscat} and \ref{tab:corr} reflect the present
understanding and the limitations due to the different choices made by
the experiments when evaluating the individual uncertainty sources.
In this preliminary result, the effects of the approximations are
evaluated by performing stability cross checks, in which the input
assumptions are changed with respect to the values reported in
Section~\ref{sec:syst}. 
The results of these cross checks are
described in the following, and summarised in Figure~\ref{fig:checks}.

\subsection{Overall correlations}
\label{sec:overallcorr}
The stability of the combined \mt\ result with respect to the
correlation assumptions reported in Table~\ref{tab:corr} has been
checked by changing, simultaneously for all systematic sources, the
values of $\rho_{\rm EXP}$, $\rho_{\rm LHC}$, $\rho_{\rm TEV}$ and $\rho_{\rm COL}$ by
a multiplicative factor, $f$, in the range [0, 1] (referred to as
$\rho_{\rm ALL}$ in the following Figures). The result of this
stability check in terms of the shifts of the combined \mt\ value
($\Delta \mt$) and of its total uncertainty ($\Delta \sigma_{\mt}$)
are reported in Figure~\ref{fig:checks}(a,b). 
While the correlated variation of all assumptions is somewhat
arbitrary, it allows the inspection of the extreme case of a combination of
totally uncorrelated measurements ($f=0$): the result is $\Delta \mt =
-114\MeV$ and $\Delta \sigma_{\mt}=-296\MeV$.

The sensitivity of the combination to the assumed correlations between
measurements from the same experiment, across experiments, and across
colliders, has been evaluated using separate variations of
$\rho_{\rm EXP}$, $\rho_{\rm LHC}$, $\rho_{\rm TEV}$ and $\rho_{\rm COL}$,
respectively. The separate correlation variations as a function of the
value of the multiplicative factor $f$ are reported by the blue,
orange, red and the grey curve in Figure~\ref{fig:checks}(a,b),
respectively.
The largest effects on $\Delta \sigma_{\mt}$ are related to the
separate variation of $\rho_{\rm EXP}$ and $\rho_{\rm COL}$, for
which, $f=0$ results in $\Delta \mt = -178\MeV$, $\Delta
\sigma_{\mt}=-114\MeV$ and $\Delta \mt =+93\MeV$, $\Delta
\sigma_{\mt}=-97\MeV$, respectively, signalling a larger sensitivity
of the result to the intra-experiment and intra-collider correlations.

An alternative study, varying separately the $\rho_{\rm EXP}$,
$\rho_{\rm LHC}$, $\rho_{\rm TEV}$ and $\rho_{\rm COL}$ correlations
for individual uncertainty sources is reported in Appendix~\ref{app:extratests}.

\subsection{JES component correlations}
The methodologies and assumptions used to derive the jet energy corrections
and the related uncertainties are not always directly comparable
between experiments. 
As a consequence, variations of the corresponding $\rho_{\rm
  LHC},~\rho_{\rm TEV}$ and $\rho_{\rm COL}$ assumptions, have been
considered in the combination stability checks. These affected the
stdJES ($\rho=0\to 0.5$), and the flavourJES components
of the JES (from $\rho=0$ to $\rho=0.5$ or $\rho=1$).  The maximum
deviations observed with respect to the default result are: $\Delta
\mt = +80\MeV$ and $\Delta \sigma_{\mt}=+22\MeV$.

A different strategy is also followed concerning the evaluation of the
$b$-jet specific energy scale uncertainty. Within the Tevatron
experiments and ATLAS, the effects stemming from $b$-quark
fragmentation, hadronisation and underlying soft radiation (the latter
for ATLAS only) are studied using different MC event generation
models~\cite{ATLASJES2}.
On the other hand, in CMS, the \Pythia\ and \Herwig\ 
fragmentation models are used to evaluate the response variation for
different jet flavor mixtures. 
The largest differences are found for pure quark and gluon flavours.
The maximum of these differences, for pure quark flavour at low
$p_{\rm T}$ and for pure gluon flavour at high $p_{\rm T}$, is taken
as a flavor uncertainty applicable to any jet flavor or flavor
mixture~\cite{CMSJES}.
To reflect these differences in the estimate of the $b$-JES
uncertainty, $\rho_{\rm LHC}=\rho_{\rm CMS-TEV}=0.5$ is used as the default
assumption for this source of systematic uncertainty. The changes of
the combination when using $\rho_{\rm LHC}=\rho_{\rm CMS-TEV} =1.0$ are
studied as another stability test. The results of this are
$\Delta \mt = -4\MeV$, $\Delta \sigma_{\mt} =  +2\MeV$.

\subsection{Signal modelling}

\subsubsection{Experimental and collider correlations}
For the evaluation of the MC systematic uncertainty, different MC
generators are used within the various collaborations
(Section~\ref{sec:calib} and Table~\ref{tab:MCsetup}). In addition a
contribution to the uncertainty due to the choice of the hadronisation
model used in the simulation is included for the Tevatron and the
ATLAS input measurements.
Finally, different input PDF (CTEQ5L and CTEQ6L1 for CDF and D0,
and CT10 and CTEQ6.6L for ATLAS and CMS, respectively) are used in
the baseline MC by the various collaborations.
These aspects may reduce the actual correlation between input
measurements for these uncertainty classes.
As a result, the combination has been repeated using
$\rho_{\rm LHC}=\rho_{\rm TEV}=\rho_{\rm COL}=0.5$ for the MC and CR uncertainty
sources: the maximum observed deviations with respect to the default
result are $\Delta \mt = +76\MeV$, and $\Delta \sigma_{\mt}= -30\MeV$
and correspond to the variation of the correlation for the signal MC
uncertainty.
In addition, variation of $\rho_{\rm COL}$ (from 0.5 to 0 or 1.0) for the
systematic uncertainties associated to the choice of the proton
(anti-proton) PDF and the modelling of the QCD radiation effects
have been considered. The results of this test are $|\Delta \mt| \le
+30\MeV$, $|\Delta \sigma_{\mt}|\le +17\MeV$.

\subsubsection{Hadronisation and alternative uncertainty categorisation}
As mentioned above, in the signal modelling categorisation, additional
uncertainties can arise from the choice of the hadronisation model
(cluster or string fragmentation as implemented in \Herwig\ and \Pythia,
respectively) describing the transition from final state partons to
colourless hadrons. The change in \mt\ obtained by exchanging cluster
and string models in a fixed MC setup can be quoted as a hadronisation
uncertainty for the \mt\ measurements.
However, this source of uncertainty is typically also considered among
the components of the jet energy scale uncertainty (both for
inclusive- and $b$-quark jets) and sizable double counting effects may
result.  For the time being, the experiments choose different
approaches. Tevatron experiments and ATLAS quote an explicit
hadronisation systematic related to the \ttbar\ modelling in the MC.
Within CMS, to minimise double counting, no additional hadronisation
systematic is quoted.
Given the relatively large size of this uncertainty (ranging between
0.27 and 0.58 \GeV\ depending on the analysis), a harmonisation of the
treatment of this systematics is needed in the future.  Specifically,
an in-depth investigation of the level of the double counting effects
involved when both types of components are used is important for the
next generation of measurements and \mt combinations. These studies
are currently in progress.
To estimate the possible significance of these effects, the \mt\
combination has been repeated for several different assumptions.  From
the comparison of \Powheg\ simulations with \Pythia\ and \Herwig\ used
for the fragmentation stage, CMS has derived estimates of the
hadronisation uncertainty of 0.58, 0.76 and 0.93 \GeV\ for the \ljets,
\dil, and \alljets channels, respectively~\cite{LHC2013}.  Adding
these into the corresponding MC systematic uncertainty, and repeating
the combination results in $\Delta \mt = -36\MeV$ and $\Delta
\sigma_{\mt}=+83\MeV$. The relatively large effect is introduced by an
increased total uncertainty for the CMS input measurements, and the
consequent change of the BLUE combination coefficients of the input
measurements.  In this case and for the eleven input measurements,
yielding: $41.5\%,~ -4.7\%,~ 7.0\%, ~8.2\%$ for the CDF \ljets,
\dil, \alljets, and \met\ measurements; $12.7\%,~0.9\%,
~20.1\%, ~-5.5\%$, for the D0 and ATLAS \ljets and \dil measurements;
and finally $~22.5\%, ~1.5\%$, and $4.1\%$ for the CMS \ljets, \dil,
and \alljets\ measurements respectively\footnote{For the eleven input
  measurements the corresponding values, in \%, of the intrinsic information
  weights are: 57.3, 5.3, 17.7, 20.8
  for CDF, 31.8, 9.2 for D0, 29.8 27.0 for ATLAS, 48.7,
  24.7, and 25.0 for CMS. The corresponding IIW$_{\rm corr}$  is $-$197.2. }.

On the other hand, if the extra hadronisation systematics evaluated by
CDF, D0, and ATLAS in addition to the JES components, are removed,
the observed changes are $\Delta \mt = +87\MeV$ and
$\Delta\sigma_{\mt}=-109\MeV$.

In addition to the above investigations, CMS has studied an
alternative systematic categorisation. While keeping the hadronisation
uncertainties described above, the bJES uncertainty is evaluated at
the analysis level using the uncertainties in the $b$-fragmentation
function, and the $b$-semileptonic branching fractions. The
uncertainty in the $b$-fragmentation is evaluated by varying the
Bowler-Lund parameters used to model the $b$-quark fragmentation in
\Pythia\ between the \Pythia\ Z2 tune and the results of the
Perugia2011~\cite{MCATLASTUNE} and Corcella ~\cite{HERWIG} tunes. This
results in an uncertainty of \mt\ of 0.15 \GeV. An additional
uncertainty of 0.10 \GeV\ comes from varying the $b$-semileptonic
branching fractions within their measured uncertainties.
In this framework, the combined uncertainty of 0.18
\GeV\ is taken as the bJES uncertainty for all CMS
input measurements. The impact of changing to this characterisation of
the hadronisation and bJES uncertainties for the CMS analyses is found
to be $\Delta \mt = -71\MeV$ and $\Delta \sigma_{\mt}= +65\MeV$.
Further work is needed to resolve this issue and detailed studies are
ongoing.

Due to the sensitivity of the combined result to the treatment of
hadronisation uncertainties, progress on these aspects will be of key
importance for future analyses of increased precision, and for
\mt\ combination updates.

\subsection{Detector modelling correlations}

The detector modelling, lepton related and $b$-tagging based
systematics could include some level of correlation between
experiments introduced by the use of MC simulation in the evaluation
of the detector performance.  For this reason, a test is performed
increasing the default correlations for these three uncertainty
sources ($\rho_{\rm TEV}, ~\rho_{\rm LHC},~\rho_{\rm COL}$) from 0\% to 50\%. The
effect of this change is found to be $\Delta \mt = -2\MeV$ and
$\Delta \sigma_{\mt}= +9\MeV$.

\subsection{Minimisation of the Fisher information }

As an additional cross check, the stability of the combination has
been verified applying the recipes described in Ref.~\cite{BLUEFIN}.
Numerical minimisation procedures aimed at reducing the Fisher
information (recall $I = 1/\sigma_{\mt}^2$) of the combination are applied varying the
correlation assumptions by multiplicative factors in three different
scenarios. In the simplest case, all correlations are rescaled by the
same global factor (minimise by global factor). As a second option,
the same rescaling factor is applied to all correlations within each
error source (minimise by error source). Finally, an alternative
minimisation procedure is performed in which for all error sources the
off-diagonal correlations ($\rho_{ij},~i\neq j$) are rescaled by the
same factor (minimise by off-diagonal element). The maximum deviations
with respect to default results are obtained for the third scenario,
and correspond to $\Delta \mt = -60\MeV$ and $\Delta \sigma_{\mt}=
+20\MeV$.

Alternative cross checks, as proposed in Ref.~\cite{BLUEFIN} and
adopted in Ref.~\cite{CMS2013}, have been performed and yield
consistent results with respect to the default combination.

\subsection{Concluding remarks on combination stability checks}

As described in the previous sub-sections, several tests varying
simultaneously the correlation assumptions for all systematic
uncertainties have been performed, changing $\rho_{\rm ALL}$, as well
as just $\rho_{\rm EXP}$, $\rho_{\rm TEV}$, $\rho_{\rm LHC}$ or $\rho_{\rm
  COL}$.
While setting $\rho_{\rm ALL} = 0$ ($f=0$) allows the inspection of the ideal
case of a combination of totally uncorrelated measurements, the
simultaneous reduction of the correlations by a factor of $1/2$ or
$4/5$ ($f=50\%,~{\rm or} ~80\%$), induces changes of the central
\mt\ value and of its total uncertainty at the level of $\pm 100\MeV$
or $\pm 40\MeV$, respectively (see Figures~\ref{fig:checkA} and
\ref{fig:checkB}).  The effect of the separate variation of the
correlations for individual uncertainty sources has been studied
(Appendix~\ref{app:extratests}), and found to be consistent (and reduced)
relative to the above results.
The stability of the world \mt\ combination under
variations of the default correlation assumptions for selected
uncertainty sources has also been studied. The largest effects are
related to changes of the correlation assumptions for the dominant
uncertainty sources: the stdJES ($\rho_{\rm TEV},~\rho_{\rm
  LHC},~\rho_{\rm COL}=0 \to 0.5$), and the MC systematic
uncertainties ($\rho_{\rm TEV},~\rho_{\rm LHC},~\rho_{\rm COL}=1 \to
0.5$). These result in $\Delta \mt = +80\MeV$, $\Delta
\sigma_{\mt}=+22\MeV$ and $\Delta \mt = +76\MeV$, $\Delta \sigma_{\mt}=
-30\MeV$, respectively (see Figures~\ref{fig:checkC} and
\ref{fig:checkD}).
In addition, the sensitivity of the combined result to the different
treatments of hadronisation uncertainties across experiments, has been
studied and estimated at the level of $\pm 100~\MeV$ for both $\Delta \mt$
and $\Delta\sigma_{\mt}$. The effect is connected to sizable changes
of the BLUE combination coefficients of the input measurements. 

Due to the relatively small size of the effects relative to the
current \mt\ precision, no additional systematic
uncertainty is associated to the final combined result.
Discussions of the methodologies used to evaluate systematic
uncertainties and the possibility to determine directly the
correlations among individual measurements are ongoing. These are
aimed at an improved treatment of the systematic uncertainties
across the experiments, and of their 
combination.

\section{Conclusions}
\label{sec:summary}

A world combination of the top-quark mass measurements
from the Tevatron and the LHC experiments has been presented.  The
result includes six measurements from Tevatron Run II and
five from the 2011 run of the LHC.

The resulting combination, taking account of
statistical and systematic uncertainties and their correlations, yields:

$$ \mt = 173.34 \pm 0.27 \mbox{ (stat)} \pm 0.71 \mbox{
  (syst)}~\GeV,$$

\noindent or, separating out the iJES statistical contribution
from the quoted systematic uncertainty:
  
$$ \mt = 173.34 \pm 0.27 \mbox{ (stat)} \pm 0.24 \mbox{ (iJES)} \pm 0.67 \mbox{ (syst)}~\GeV.$$

The world combination achieves an improvement of the total \mt\
uncertainty of 28\% relative to the most precise single input measurement~\cite{CMSlj2011}
and $\approx 13\%$ relative to the previous most precise
combination~\cite{TEV2013}.
The total uncertainty of the combination is 0.76~\GeV, and is
currently dominated by systematic uncertainties due to jet
calibration and modelling of the \ttbar\ events.
Given the current experimental uncertainty on \mt, clarifying the
relation between the top quark mass implemented in the MC and the
formal top quark pole mass demands further theoretical investigations.
The dependence of the result on the correlation assumptions between
measurements from the same experiment and across experiments has been
studied and found to be small compared to the current \mt
precision. 
At the same time, the results of the stability tests reveal the importance of
the ongoing discussions on the methodologies used for the evaluation of
the systematic uncertainties. In some cases, experiments adopt
different approaches which deserve further studies and
harmonisation.

%% file: appendix.tex

\section{Uncertainty naming conventions}
\label{app:naming}

In this Appendix the naming conventions for each systematic uncertainty
are summarised. The convention used for the present analysis is
compared with those from Refs.~\cite{TEV2013, LHC2013,
  CDF-D0-Combo2012}  (see Table~\ref{tab:namingconv}).

\begin{table}[!th]
\begin{center}
\footnotesize 
\begin{tabular}{|r|r|rr|}
\hline

WA   & LHC comb~\cite{LHC2013} & \multicolumn{2}{|c|}{TEV comb.~\cite{TEV2013} (~\cite{CDF-D0-Combo2012})} \\
\hline
Stat       &   Statistics &   Statistics  & (Statistical uncertainty) \\ 
iJES       &   iJES       &   iJES        & ({\em in situ} light-jet calibration)\\ 
stdJES     &   uncorrJES$\oplus$insitu$\gamma/Z$JES$\oplus$intercalibJES  & dJES$\oplus$cJES$\oplus$rJES &
(Light-jet response 1$\oplus$2)$\oplus$Out-of-cone correction\\ 
flavourJES &  flavourJES    & aJES & (Response to $b/q/g$ jets)\\ 
bJES       &  bJES  &  bJES & (Model for $b$ jets)\\ 
\hline
MC         &  MC$\oplus$UE  & part of Signal& (part of Signal modelling)  \\ 
Rad        &  Rad  & part of Signal& (part of Signal modelling)  \\ 
CR         &  CR   & part of Signal& (part of Signal modelling)  \\ 
PDF        &  PDF  & part of Signal& (part of Signal modelling)  \\ 
\hline
DetMod     &  DetMod  & DetMod &(Jet modelling) \\ 
$b$-tag    &  $b$-tagging  & part of BGData & (part of background based on data)\\ 
LepPt      &  Lepton reconstruction  &  LepPt & (Lepton modelling)\\ 
BGMC       &  Background from MC    & BGMC & (Background from theory)\\ 
BGData       &  Background from Data  & BGData & (Background based on data)\\ 
Meth       &  Method  & Method & (Calibration Method) \\ 
MHI        &  Multiple Hadronic Interactions  & MHI & (Multiple
interaction model) \\ 
\hline
\end{tabular}
\caption{Uncertainty naming conventions used for this combination, and
from Refs.~\cite{TEV2013, LHC2013, CDF-D0-Combo2012}. }
\label{tab:namingconv}

\end{center}
\end{table}


\section{Additional figures}
\label{app:extrafigs}
 
 In this appendix, the result of the standard combination is compared
 with those from the correlated \mt\ determination per \ttbar\ decay
 channel, experiments and collider. With respect to
 Figure~\ref{fig:sepcomb}, the \mt\ parameters are influenced by the
 full set of input measurements through their correlations.

 \begin{figure}[t!]
 \begin{center}
 \includegraphics[width=0.77\textwidth]{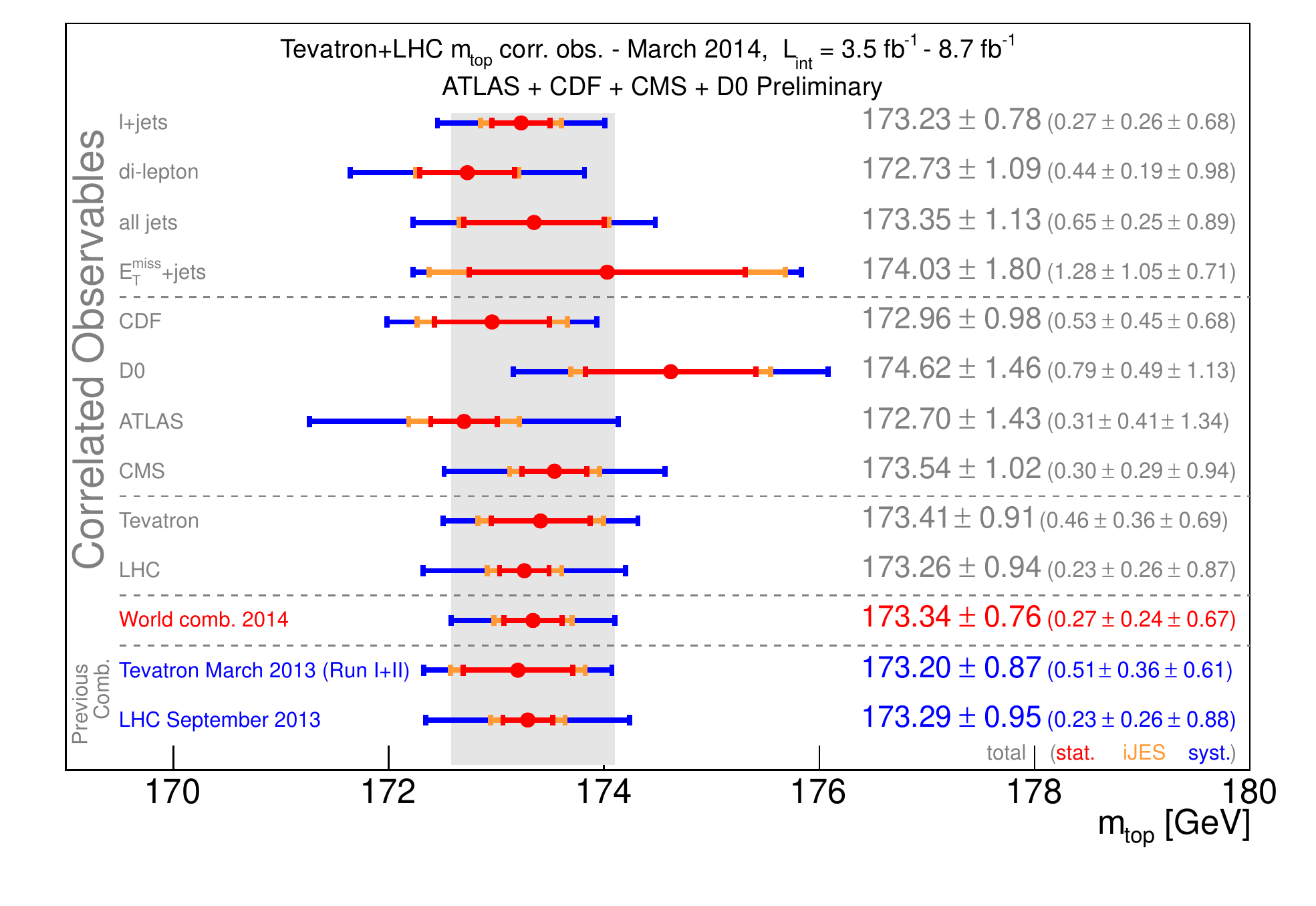}
 \end{center}
 \caption{Comparison of the standard combination results with the
   correlated \mt\ determinations (parameter values in
   Tables~\ref{tab:channels}, \ref{tab:exp} and \ref{tab:col}) per
   \ttbar\ decay channel, experiment, and collider. The grey vertical
   band reflect the total uncertainty on the combined \mt\ value.}
 \label{fig:corrcomb}

 \begin{center}
 \includegraphics[width=0.77\textwidth]{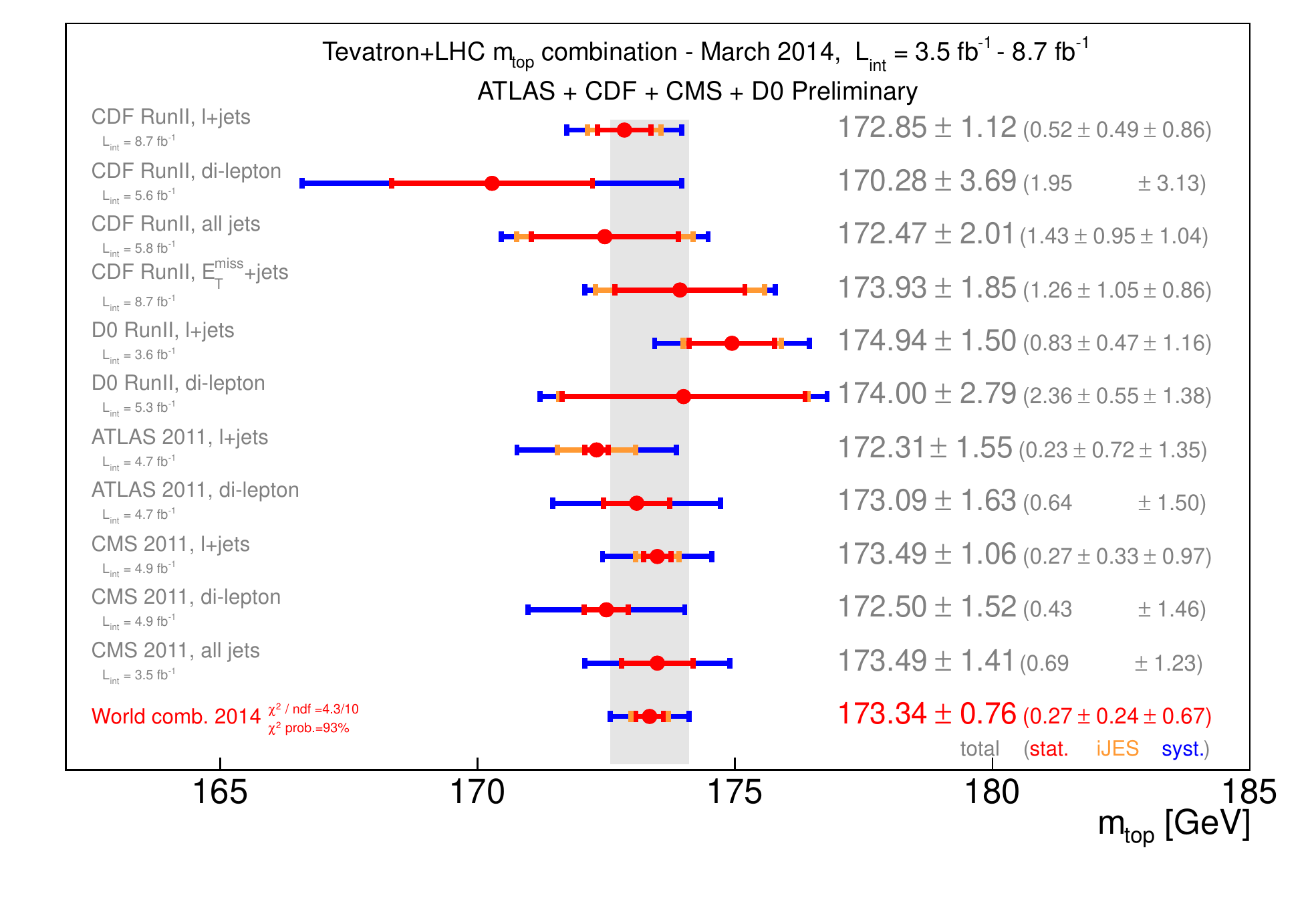}
 \end{center}
 \caption{Input measurements and result of the world combination (see
   also Table~\ref{tab:syscat}).  For each measurement, the total
   uncertainty, the statistical and the iJES contributions (when
   applicable) as well as the sum of the remaining uncertainties are
   reported separately. The iJES contribution is statistical in nature
   and applies to analyses performing in situ (\ttbar) jet energy
   calibration procedures.  The grey vertical band indicates the total
   world \mt\ uncertainty. With respect to Figures~\ref{fig:summaryA},
   the comparison with previous individual Tevatron and LHC
   combinations~\cite{TEV2013, LHC2013} is removed.}
 \label{fig:summaryAalternative}
 \end{figure}

 In addition, Figure~\ref{fig:summaryAalternative} reports an
 alternative summary plot. With respect to Figures~\ref{fig:summaryA},
 the comparison with previous individual Tevatron and LHC
 combinations~\cite{TEV2013, LHC2013}  is removed.

\clearpage

\section{Results of the individual experiment combinations}
\label{app:sepcomb}

In this Appendix the separate experiment and collider combinations are
reported and compared to the \mt\ world results
(Table~\ref{tab:sepcomb}).
The results of the individual combinations are obtained neglecting
other input measurements and their correlations.


\begin{table}[!h]
\begin{center}
\footnotesize 
\begin{tabular}{|r|c|c|c|c||c|c||c|}
\hline

All values in \GeV & CDF & D0 &  ATLAS & CMS & Tevatron & LHC & WA \\ \hline
\mt       & 173.19 & 174.85 & 172.65 & 173.58 & 173.58 & 173.28 & 173.34 \\ \hline
Stat      &   0.52 &   0.78 &   0.31 &   0.29 &   0.44 &   0.22 &   0.27 \\ \hline
iJES      &   0.44 &   0.48 &   0.41 &   0.28 &   0.36 &   0.26 &   0.24 \\ 
stdJES    &   0.30 &   0.62 &   0.78 &   0.33 &   0.27 &   0.31 &   0.20 \\ 
flavourJES&   0.08 &   0.27 &   0.21 &   0.19 &   0.09 &   0.16 &   0.12 \\ 
bJES      &   0.15 &   0.08 &   0.35 &   0.57 &   0.13 &   0.44 &   0.25 \\ \hline
MC        &   0.56 &   0.62 &   0.48 &   0.19 &   0.57 &   0.25 &   0.38 \\ 
Rad       &   0.09 &   0.26 &   0.42 &   0.28 &   0.13 &   0.32 &   0.21 \\ 
CR        &   0.21 &   0.31 &   0.31 &   0.48 &   0.23 &   0.43 &   0.31 \\ 
PDF       &   0.09 &   0.22 &   0.15 &   0.07 &   0.12 &   0.09 &   0.09 \\ \hline
DetMod    &   $<$0.01 &   0.37 &   0.22 &   0.25 &   0.09 &   0.20 &   0.10 \\ 
$b$-tag   &   0.04 &   0.09 &   0.66 &   0.11 &   0.04 &   0.22 &   0.11 \\ 
LepPt     &   $<$0.01 &   0.20 &   0.07 &   $<$0.01 &   0.05 &   0.01 &   0.02 \\ 
BGMC      &   0.10 &   0.16 &   0.06 &   0.11 &   0.11 &   0.08 &   0.10 \\ 
BGData      &   0.15 &   0.19 &   0.06 &   0.03 &   0.12 &   0.04 &   0.07 \\ 
Meth      &   0.07 &   0.15 &   0.08 &   0.07 &   0.06 &   0.06 &   0.05 \\ 
MHI       &   0.08 &   0.05 &   0.02 &   0.06 &   0.06 &   0.05 &   0.04 \\ \hline                                                       
Total Syst&   0.85 &   1.25 &   1.40 &   0.99 &   0.82 &   0.92 &   0.71 \\ 
Total     &   1.00 &   1.48 &   1.44 &   1.03 &   0.94 &   0.94 &   0.76 \\ 
\hline

$\chi^2$/ndf   &   1.09 / 3 &   0.13 / 1 &   0.34 / 1 &   1.15 / 2 &
2.45 / 5 & 1.81
/ 4 &   4.33 / 10\\ 
$\chi^2$ probability [\%]&   78 &   72 &   56 &   56 &   78 &   77 &   93 \\ 
\hline
\end{tabular}

\caption{Results of the individual experiment and collider
  combinations using the inputs listed in Table~\ref{tab:syscat}. The
  uncertainty breakdown is provided and compared with the results of
  the \mt\ world combination.}
\label{tab:sepcomb}

\end{center}
\end{table}

\clearpage

\section{Additional stability tests}
\label{app:extratests}

\begin{figure}
\begin{center}
\subfigure[$\Delta \mt$ varying $\rho_{\rm EXP}$]{\label{extrafig1:A} \includegraphics[width=0.48\textwidth]{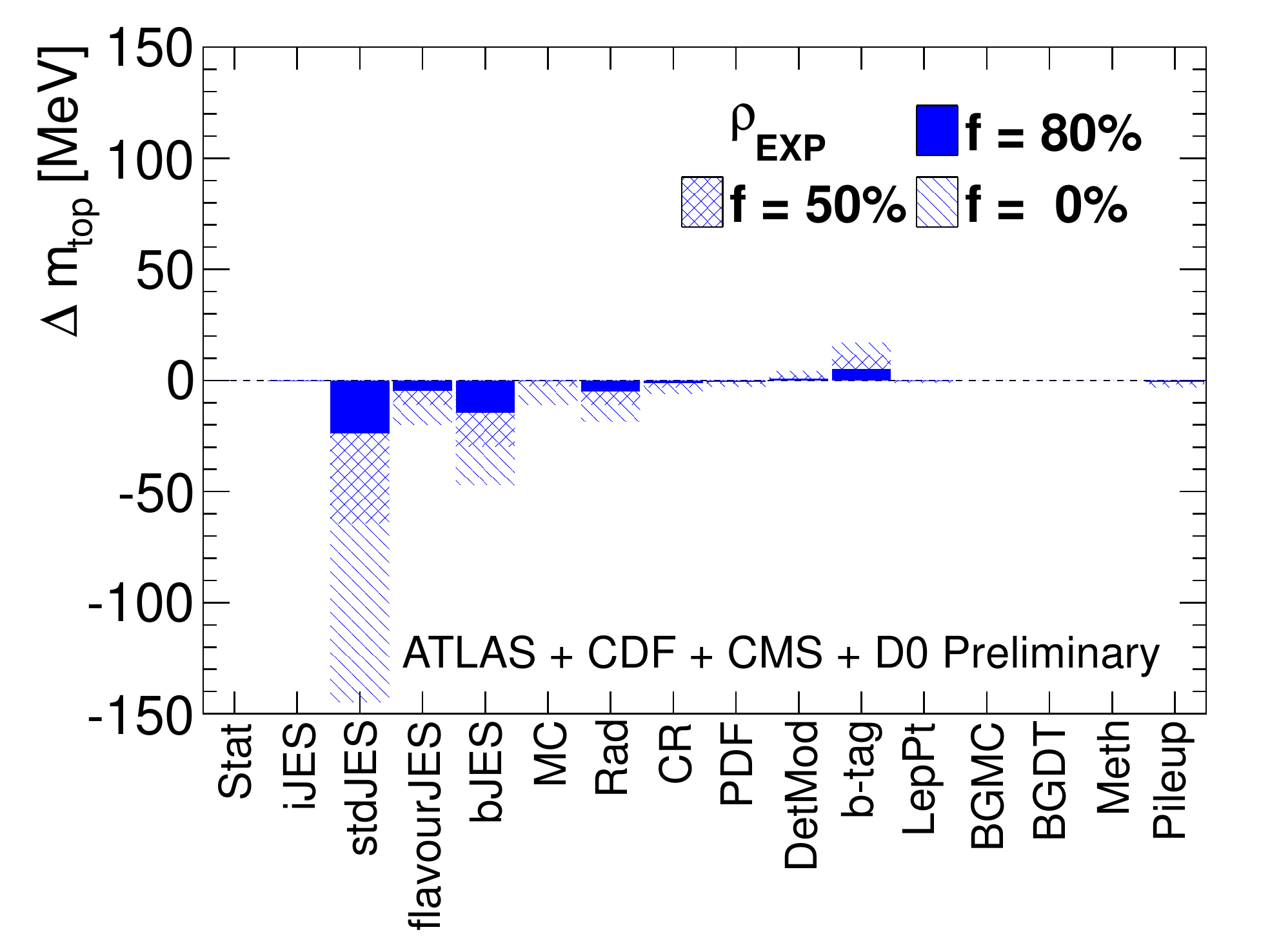}}
\subfigure[$\Delta \sigma(\mt)$ varying $\rho_{\rm EXP}$]{\label{extrafig1:B} \includegraphics[width=0.48\textwidth]{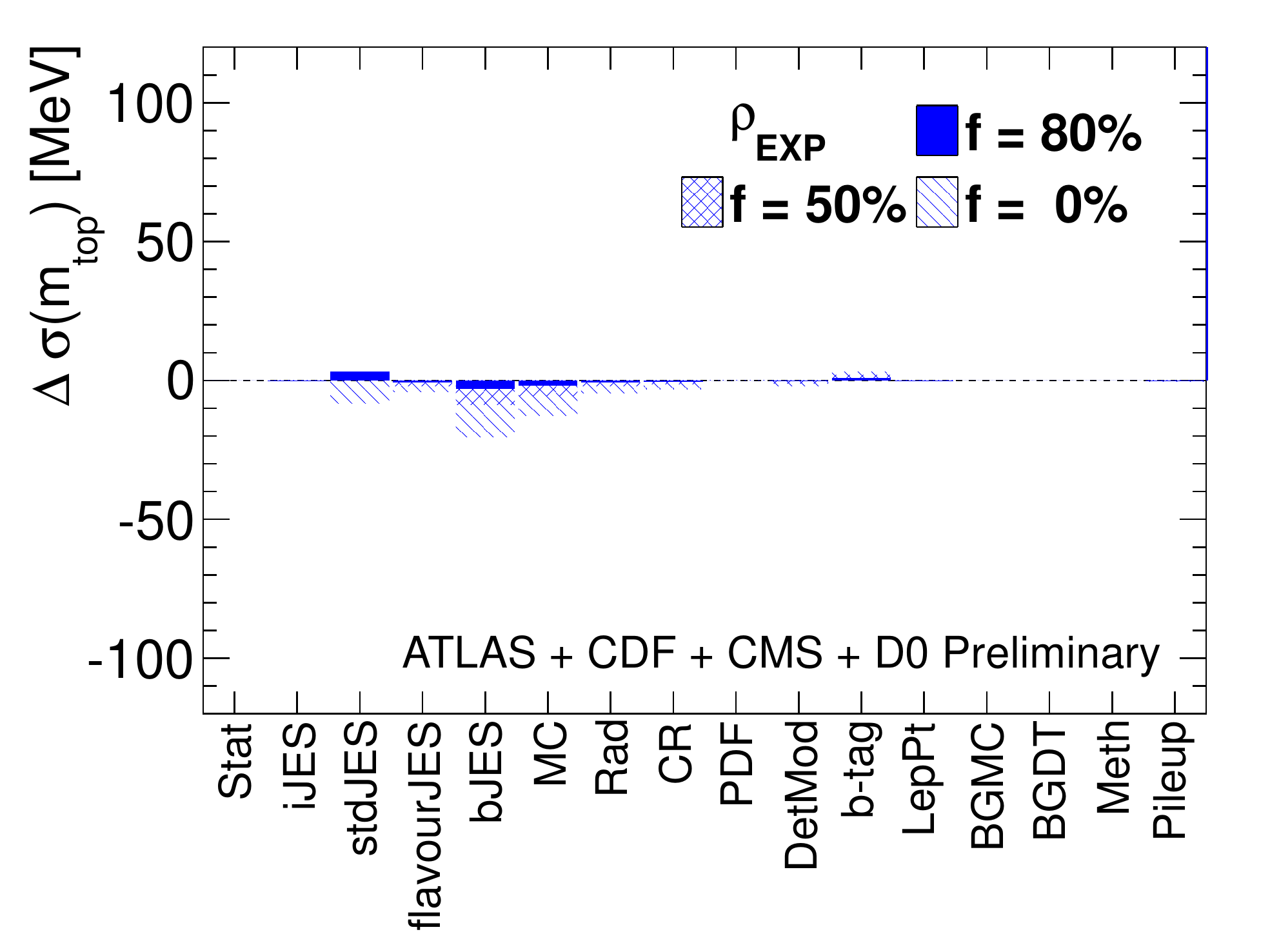}}

\subfigure[$\Delta \mt$ varying $\rho_{\rm TEV}$]{\label{extrafig1:C} \includegraphics[width=0.48\textwidth]{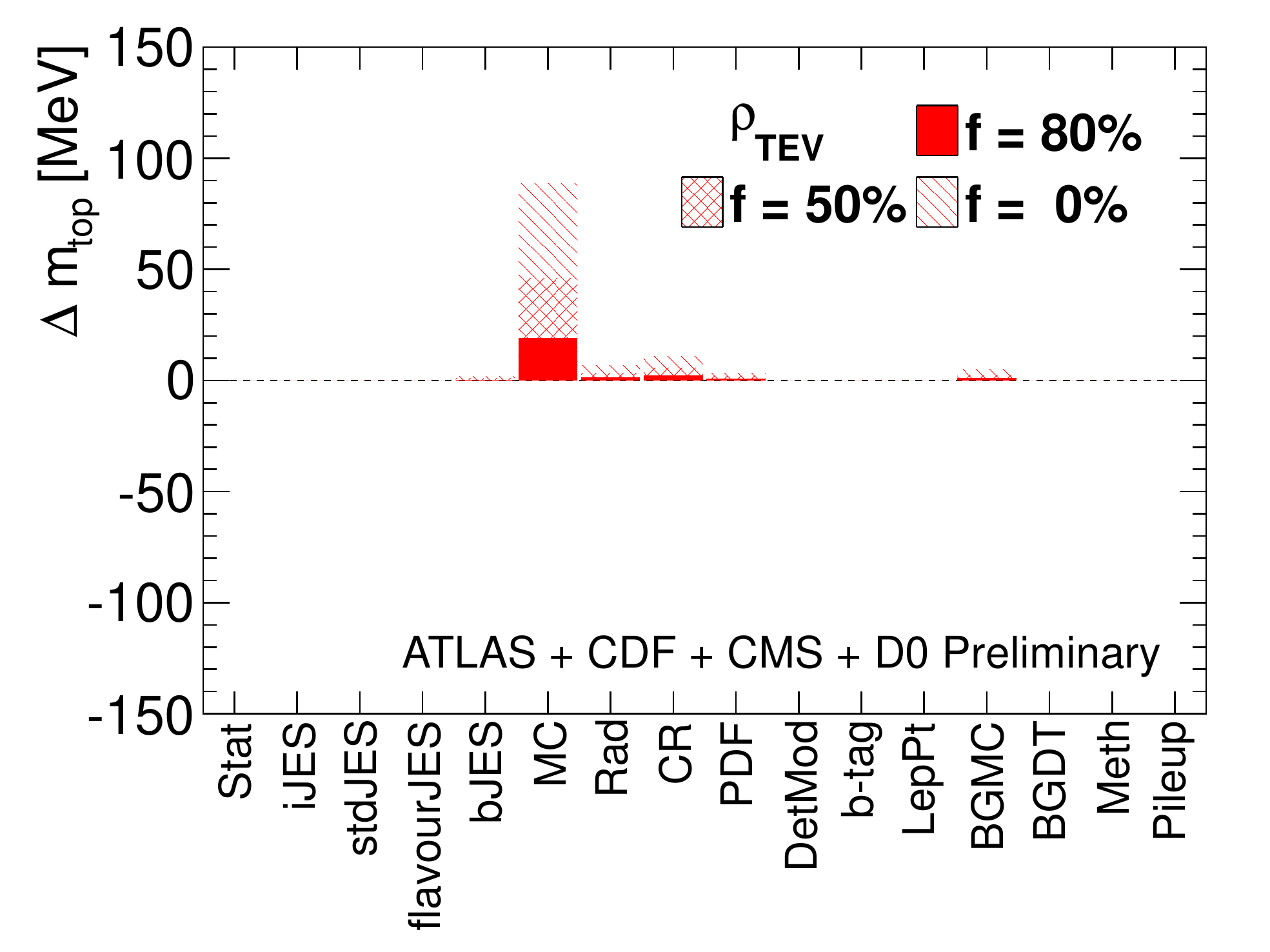}}
\subfigure[$\Delta \sigma(\mt)$ varying $\rho_{\rm TEV}$]{\label{extrafig1:D} \includegraphics[width=0.48\textwidth]{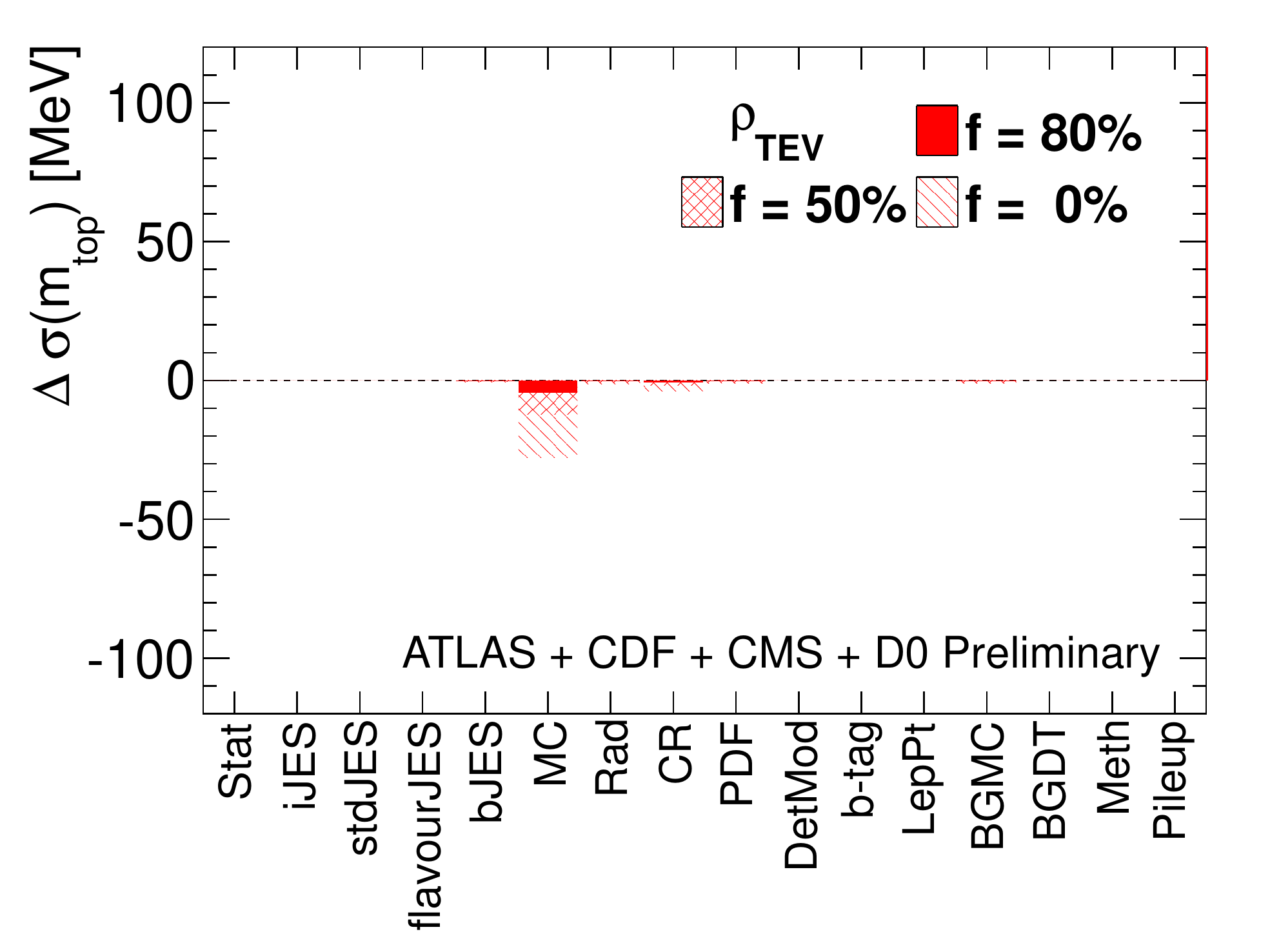}}

\subfigure[$\Delta \mt$ varying $\rho_{\rm LHC}$]{\label{extrafig1:E} \includegraphics[width=0.48\textwidth]{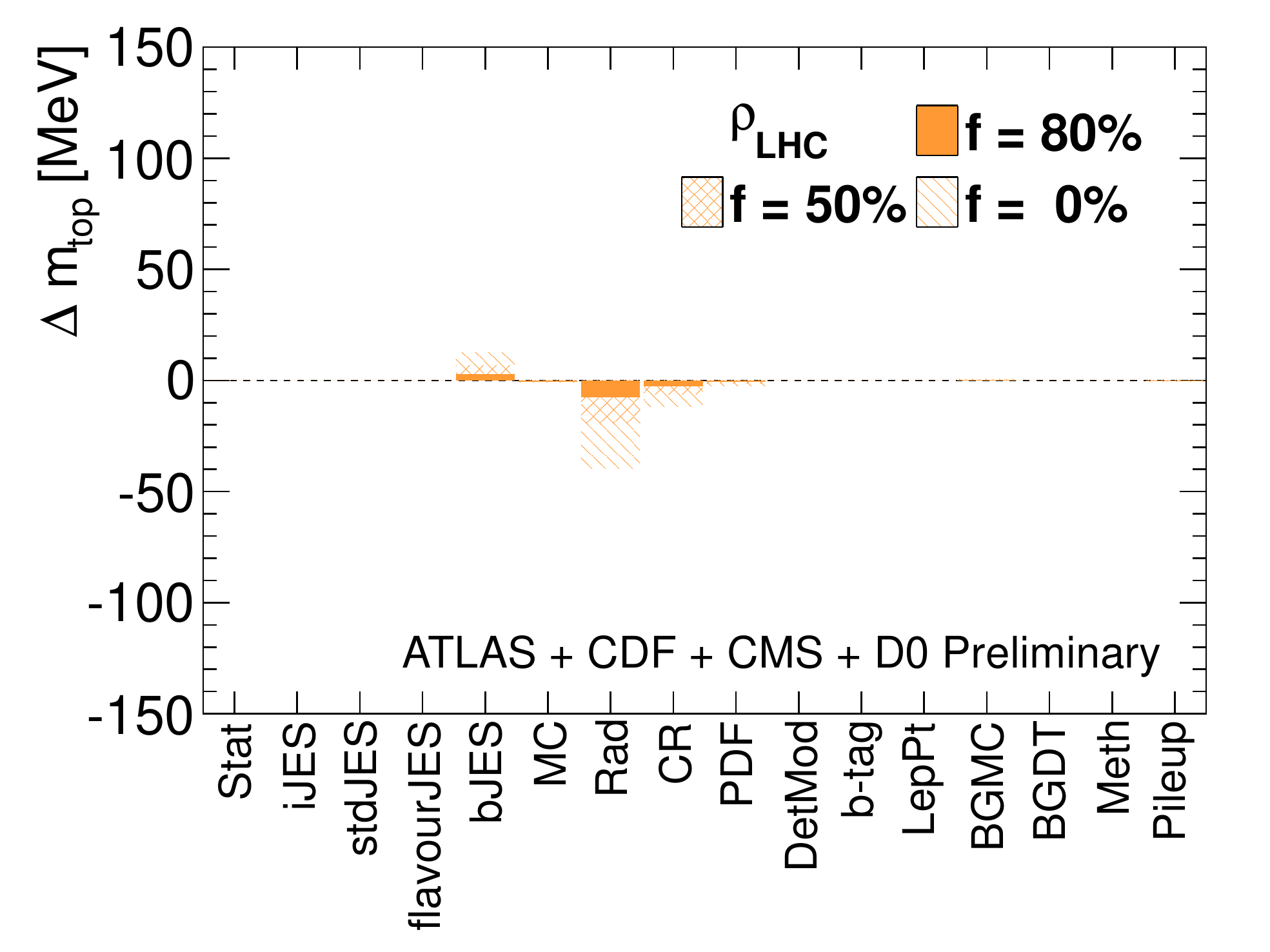}}
\subfigure[$\Delta \sigma(\mt)$ varying $\rho_{\rm LHC}$]{\label{extrafig1:F} \includegraphics[width=0.48\textwidth]{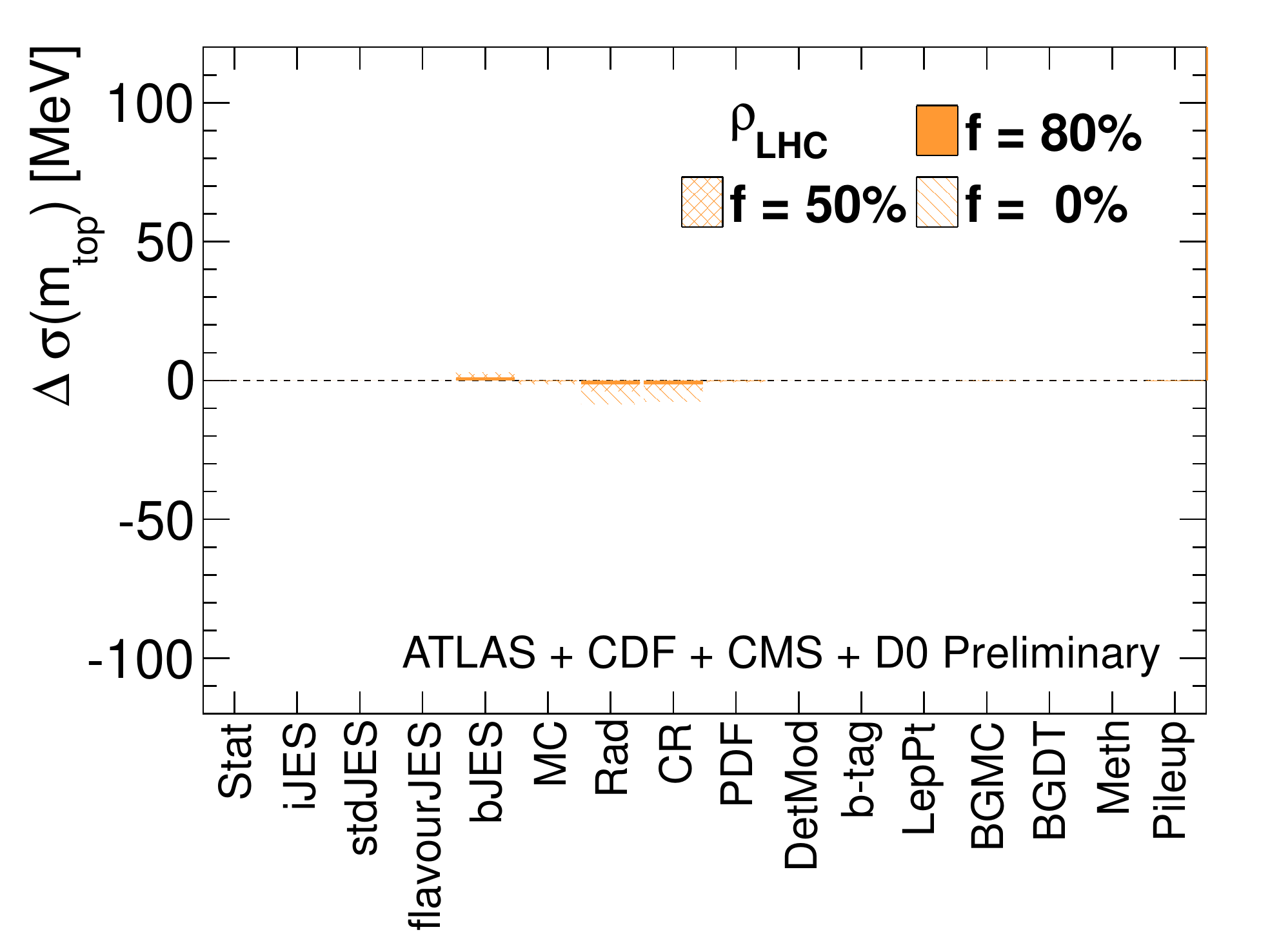}}

\end{center}
\caption{Variation of the combined \mt\ value (a, c, e) and of its
  uncertainty (b, d, f) for three different 
  correlation assumptions for each uncertainty category ($f\cdot \rho$,
  with $f=0.8,~0.5$, and $0$). Variation of $\rho_{\rm EXP}$,
  $\rho_{\rm TEV}$ and $\rho_{\rm LHC}$ are reported by the top,
  middle and bottom panels, respectively. }
\label{fig:extratest1}

\end{figure}
\begin{figure}
\begin{center}
\subfigure[$\Delta \mt$ varying $\rho_{\rm COL}$]{\label{extrafig2:A} \includegraphics[width=0.48\textwidth]{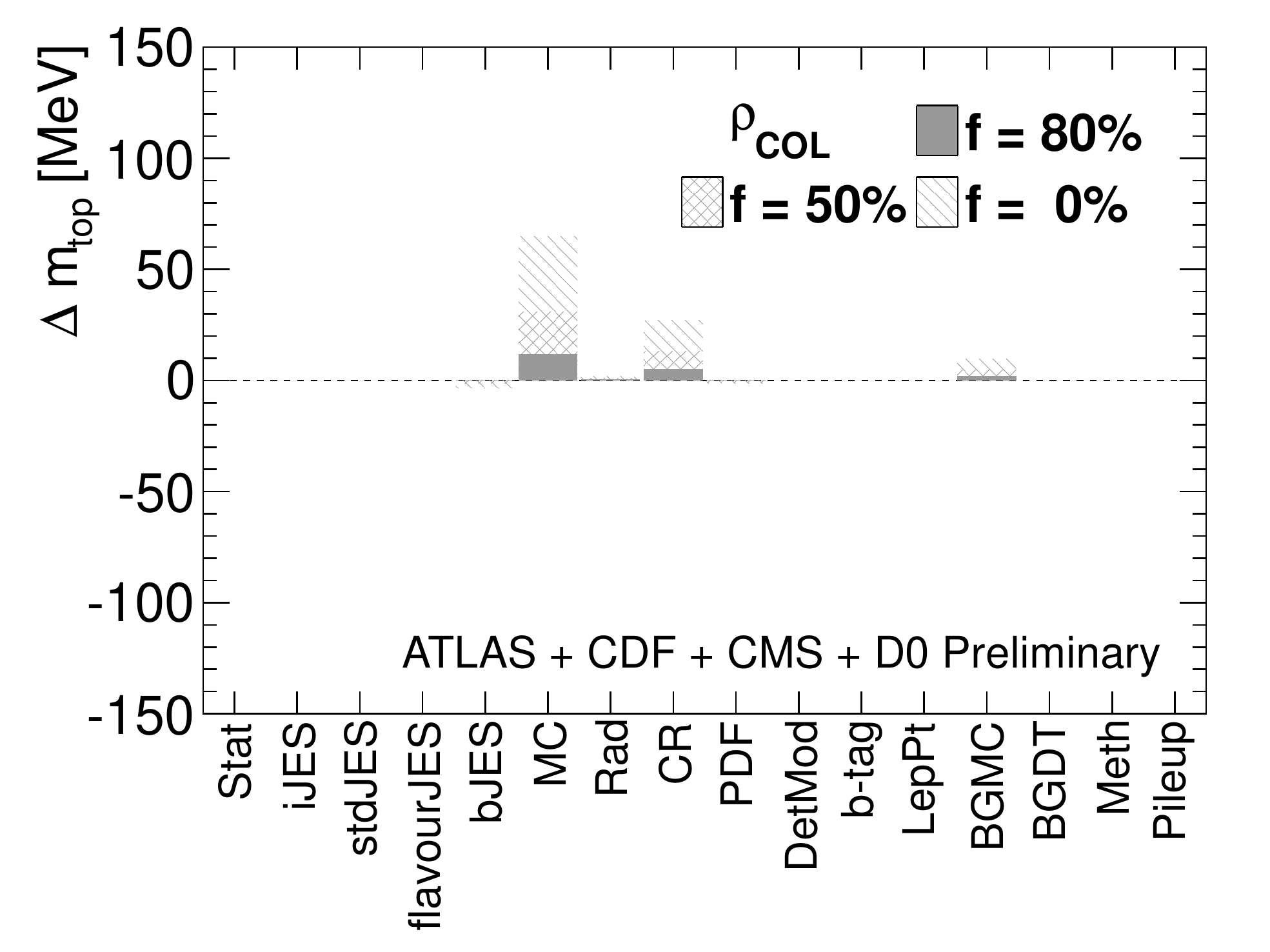}}
\subfigure[$\Delta \sigma(\mt)$ varying $\rho_{\rm COL}$]{\label{extrafig2:B} \includegraphics[width=0.48\textwidth]{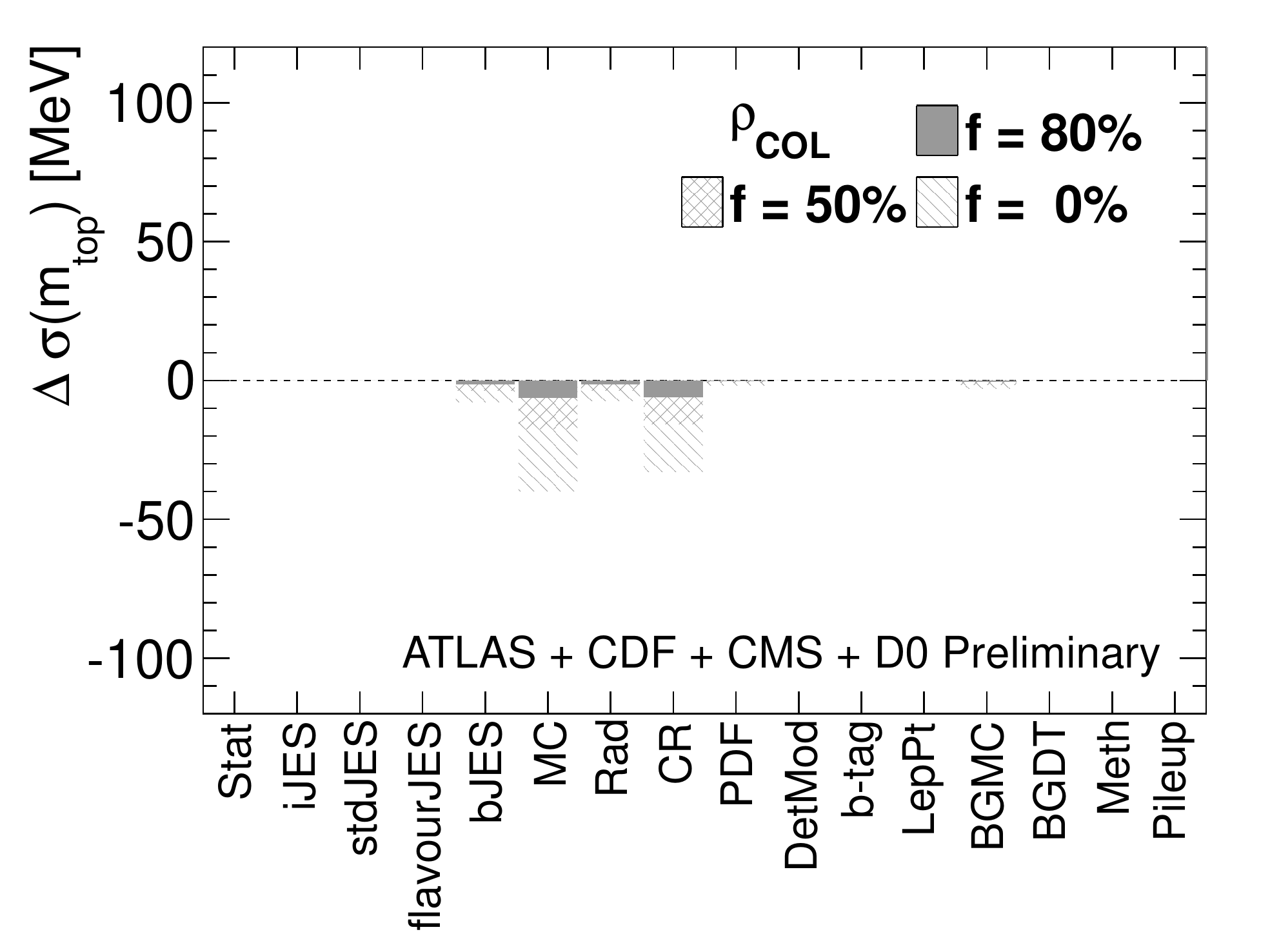}}

\subfigure[$\Delta \mt$ varying $\rho_{\rm ALL}$]{\label{extrafig2:C} \includegraphics[width=0.48\textwidth]{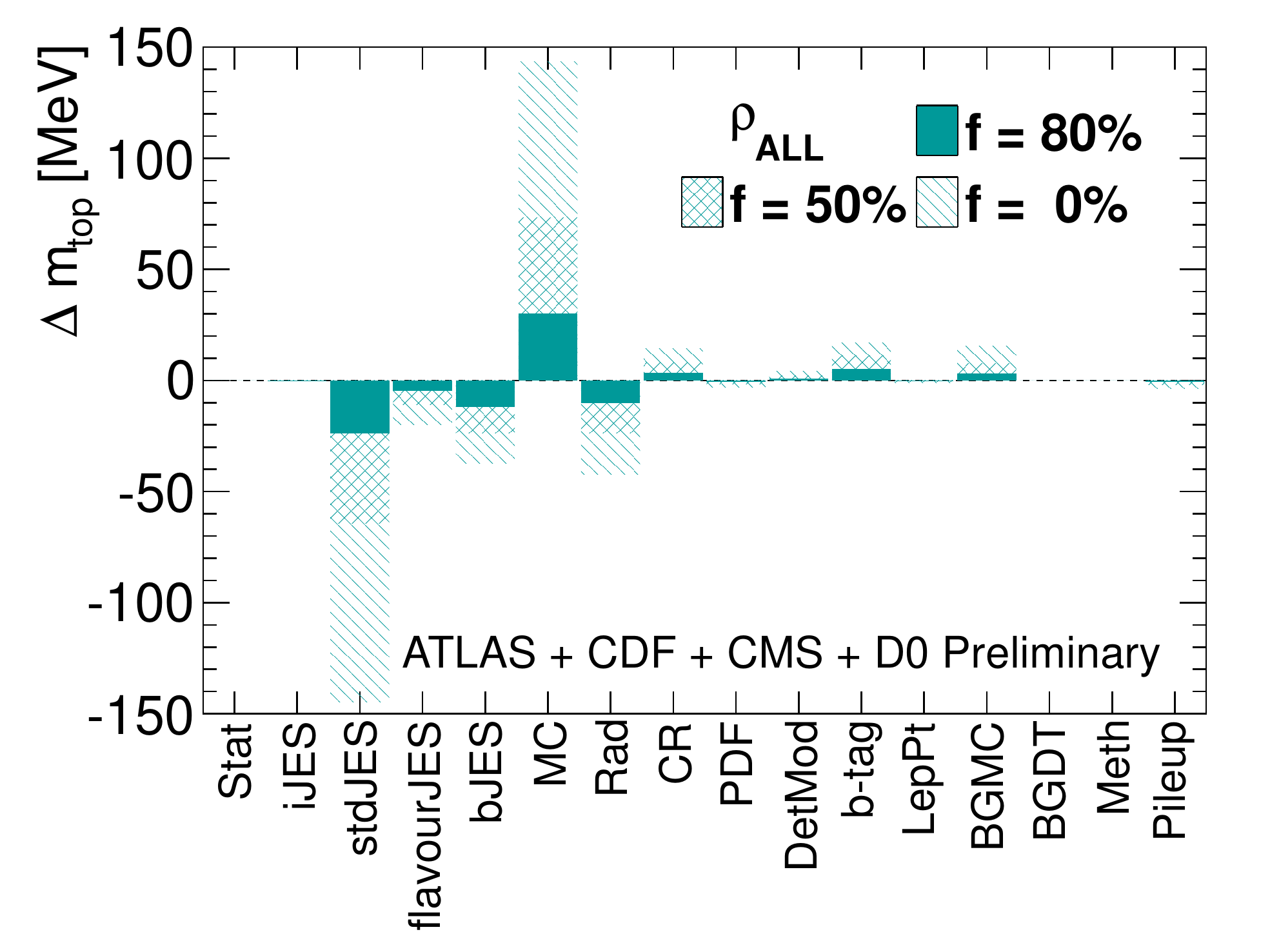}}
\subfigure[$\Delta \sigma(\mt)$ varying $\rho_{\rm ALL}$]{\label{extrafig2:D} \includegraphics[width=0.48\textwidth]{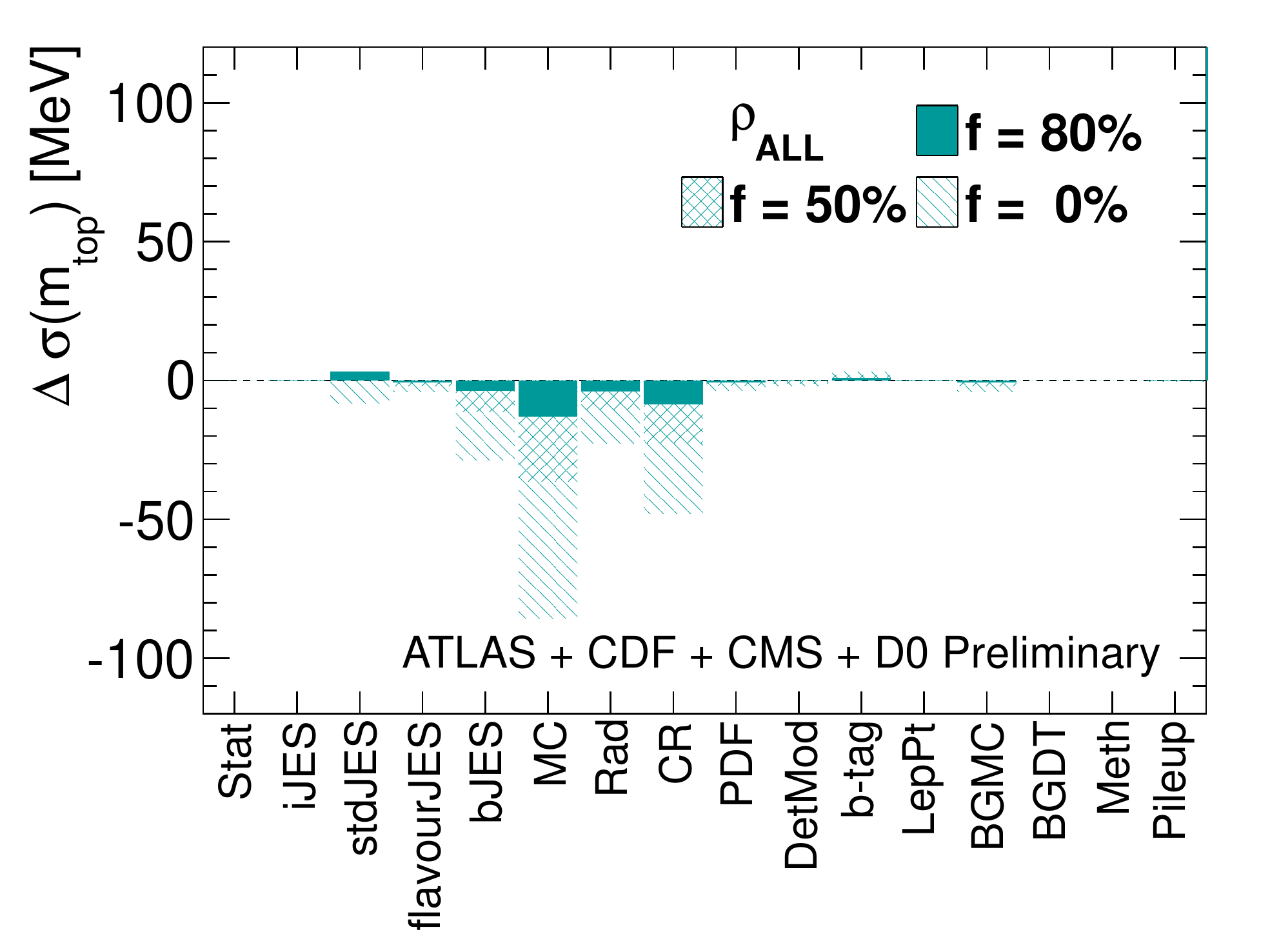}}

\end{center}
\caption{Variation of the combined \mt\ value (a, c) and of its
  uncertainty (b, d) for three different
  correlation assumptions for each uncertainty category ($f\cdot \rho$,
  with $f=0.8,~0.5$, and $0$). Variation of $\rho_{\rm COL}$, and for
  all correlation between uncertainty sources ($\rho_{\rm ALL}$) are 
  reported by the top, and bottom panels, respectively. }
\label{fig:extratest2}
\end{figure}

In this Appendix, the stability tests reported in
Section~\ref{sec:overallcorr}, are complemented by separately varying 
the $\rho_{\rm EXP}$, $\rho_{\rm LHC}$, $\rho_{\rm TEV}$ and
$\rho_{\rm COL}$ correlations for the individual uncertainty sources~\cite{BLUERN}.
This study rests on the assumptions that all uncertainty classes are
uncorrelated with respect to each other, and allow the identification
of those uncertainty categories for which the correct assessment of the
correlation is important for the stability of the result.

In Figures~\ref{fig:extratest1} and \ref{fig:extratest2} the variation
of the combined \mt\ value (left panels) and of its uncertainty (right
panels) are reported for three different variations of the
correlation assumptions for each of the uncertainty categories ($f\cdot
\rho$, with $f=0.8,~0.5$, and $0$). 
Figure~\ref{fig:extratest1} reports the investigations for $\rho_{\rm
  EXP}$ (top panels), $\rho_{\rm LHC}$ (middle panels), and $\rho_{\rm
  TEV}$ (bottom panels). Figure~\ref{fig:extratest2} displays the
results for the variation of $\rho_{\rm COL}$ (top panels), and for
the simultaneous variations of all correlation assumptions, $\rho_{\rm
  ALL}$ (bottom panels).

The largest observed variations of the combined \mt\
result are of the order of about 150~\MeV\ (for $f=0$), and are 
related to changes of the correlation assumptions of the JES
uncertainty categories (Figure~\ref{extrafig1:A}, for
$\rho_{\rm EXP}$), the MC (Figure~\ref{extrafig1:C}, and
Figure~\ref{extrafig2:A} for $\rho_{\rm TEV}$ and $\rho_{\rm
  COL}$, respectively), the Radiation (Figure~\ref{extrafig1:E}, for
$\rho_{\rm LHC}$), and the CR (Figure~\ref{extrafig2:A}, for
$\rho_{\rm COL}$) systematics, respectively.  As expected, a
combination of the above effects is observed when varying all
correlation assumptions between input measurements ($\rho_{\rm ALL}$),
regardless the experiment or collider they originate from
(Figure~\ref{extrafig2:C}).

The variation of the total combined uncertainties are typically more
contained ($\Delta(\sigma(\mt))< 100~\MeV$), and negative (reducing
the correlation increases the precision of the combined result). The
sources reporting the largest sensitivities are related to the
variation of $\rho_{\rm LHC}$, $\rho_{\rm TEV}$, $\rho_{\rm COL}$, and
$\rho_{\rm ALL}$, for the JES and MC modelling uncertainties (MC,
Radiation, CR).
An exception is made concerning the stdJES and $b$-tagging systematic
categories, for which the reduction of the correlation assumption can
yield a slight increase (of the order of about $ 5\MeV$) of
$\Delta(\sigma(\mt))$. This is a consequence of the relatively high
correlations between the input measurements~\cite{BLUEFIN}. The effect is
however negligible relative to the present total uncertainty on \mt.

%% file: main.bbl
\begin{thebibliography}{100}
\bibitem{Z2TUNE}
 {\rm The Z2 tune is identical to the Z1 tune described in \cite{Z1TUNE} except
  that Z2 uses the CTEQ6L PDF while Z1 uses CTEQ5L}.
\end{thebibliography}
